\documentclass[10pt,final,journal,twocolumn]{IEEEtran}
\ifCLASSINFOpdf

\else
\usepackage{algorithm}
\usepackage{booktabs}

\fi
\usepackage{cite}
\usepackage[cmex10]{amsmath}
\usepackage{amsthm}
\usepackage{stfloats}
\usepackage{multicol,multienum}
\usepackage{multirow}
\usepackage{mdframed}
\usepackage{amssymb}
\usepackage{url}
\usepackage{amsmath}
\usepackage{xcolor}
\usepackage[dvips]{graphicx}
\usepackage{subfigure}
\usepackage{caption}
\usepackage{amsmath}
\usepackage{amsfonts,amssymb}
\usepackage{bbm}
\usepackage[T1]{fontenc}
\usepackage{algorithm}
\usepackage{algpseudocode}
\usepackage{ulem}
\usepackage{balance}
\usepackage[linesnumbered, ruled]{}
\makeatletter
\renewcommand{\maketag@@@}[1]{\hbox{\m@th\normalsize\normalfont#1}}%
\makeatother

\usepackage{caption}
\begin{document}
\hyphenation{op-tical net-works semi-conduc-tor}
\title{Fundamental Tradeoff in Movable Antenna Systems: How Long to Move Before Transmission?}
\author{Guojie Hu, Qingqing Wu, Lipeng Zhu, Wen Chen, and Shanpu Shen\vspace{-1.2em}
\thanks{
%
G. Hu is with the College of Communication Engineering, Rocket Force University of Engineering, Xi'an 710025, China (email: lgdxhgj@sina.com). Q. Wu and W. Chen are with the Department of Electronic Engineering, Shanghai Jiao Tong University, Shanghai 200240, China (email: qingqingwu@sjtu.edu.cn; wenchen@sjtu.edu.cn). L. Zhu is with the State Key Laboratory of CNS/ATM and the School of Interdisciplinary Science, Beijing Institute of Technology, Beijing 100081, China (e-mail: zhulp@bit.edu.cn). S. Shen is with the State Key Laboratory of Internet of Things for Smart City and Department of Electrical and Computer Engineering, University of Macau, Macau, China (e-mail: shanpushen@um.edu.mo).
}
}
\IEEEpeerreviewmaketitle
\maketitle
\begin{abstract}
The movable antenna (MA) technology enables flexible reconfiguration of wireless channels through adaptive antenna deployment, offering significant potential for enhancing communication performance. However, antenna movement requires a certain duration within which communication may be compromised due to factors such as channel fluctuation and Doppler effect. This leads to a fundamental tradeoff: A longer movement duration allows antennas to reach more favorable positions for better channel conditions, but it inevitably reduces the time available for data transmission. To characterize the aforementioned tradeoff, we focus on the MAs-enabled multiuser downlink scenario, and jointly optimize the movement duration and antenna deployment at the base station to maximize the effective throughput. The formulated problem is highly non-convex. The general solutions require an one-dimensional search over movement durations, each with optimized antenna deployment. To reduce complexity, we propose a fitting method that samples only a few rate-duration pairs, yielding a closed-form expression that captures the rate trend and enables a favorable solution immediately. We further derive a closed-form condition on the maximum antenna movement speed: When the speed is below a certain threshold, the optimal strategy is to keep antennas stationary throughout the transmission period. The fundamental tradeoff and the effectiveness of the proposed solutions are examined in a special case with two MAs and two users. Finally, numerical simulations validate the efficacy of the proposed schemes.
\end{abstract}

\IEEEpeerreviewmaketitle
\vspace{-10pt}
\section{Introduction}
Multiple-input multiple-output (MIMO) technology has advanced wireless communications by leveraging spatial diversity and multiplexing gains to enhance link reliability and spectral efficiency \cite{MIMO_1,MIMO_2}. However, conventional MIMO architectures predominantly employ fixed-position antennas (FPAs), whose static spatial configuration limits adaptability to fluctuating wireless environments  and user distributions \cite{MIMO_3}. Under deep fading or strong spatial correlation, the benefits of beamforming and spatial multiplexing degrade significantly, resulting in suboptimal hardware utilization. This motivates the development of more agile antenna frameworks capable of dynamic channel adaptation without excessive hardware overhead.

 \begin{figure*} [!t]
\centering
\includegraphics[width=15cm]{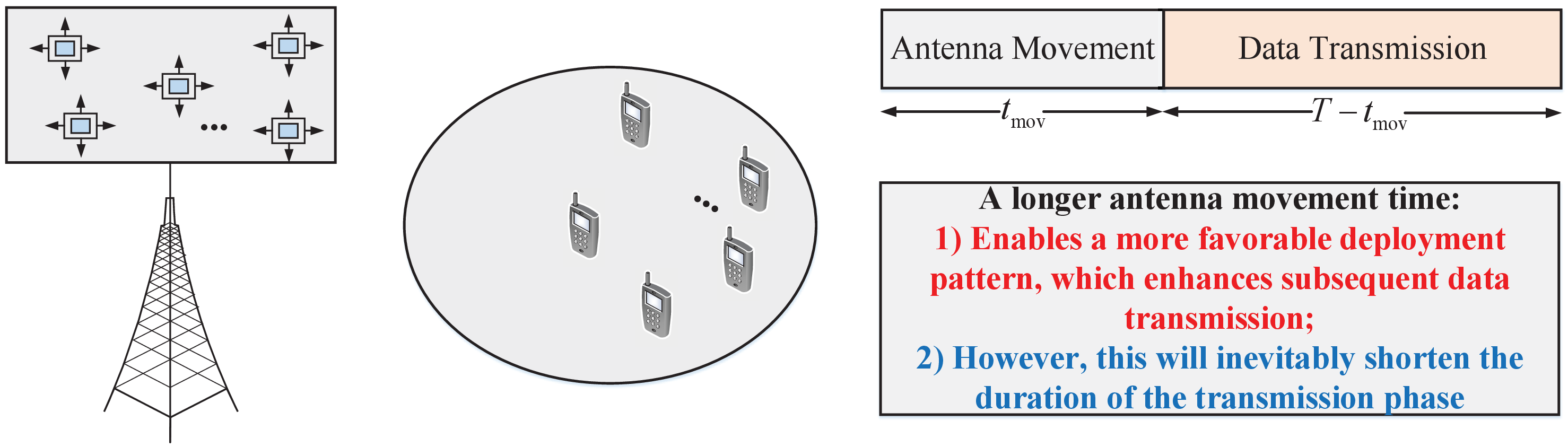}
\captionsetup{font=small}
\caption{Illustration of the considered system model and the fundamental tradeoff.} \label{fig:Fig1}
\vspace{-5pt}
\end{figure*}

Movable antenna (MA) technology has emerged as a promising solution to overcome FPA limitations \cite{Lipeng1,KKWong1,Lipeng2,Xiaodan1}. For example, by integrating antennas with mechanical actuators such as stepper motors or servos, MAs can be physically repositioned within a confined region, enabling on-demand reconfiguration of wireless channels \cite{Ningboyu,Weidong}. This introduces an additional spatial degree of freedom (DoF), allowing antennas to be deliberately deployed at positions that yield higher channel gains, reduced interference, or improved spatial channel separation among users. Consequently, MA-aided systems can achieve substantial performance gains without increasing the number of antennas and associated radio frequency (RF) chains, offering a cost-effective path toward more adaptive and efficient next-generation networks \cite{Xiaodan2}.

The considerable promise of MAs has spurred extensive investigations across a broad perspective of wireless communication contexts. Early contributions laid the groundwork for channel characterization, notably through the field-response modeling approach tailored for narrowband single-input single-output (SISO) links \cite{Lipeng1}, which has subsequently been expanded to accommodate wideband transmissions \cite{Lipeng3,Lipeng4}, flexible beamforming \cite{Lipeng_CL}, generic MIMO configurations \cite{Wenyan1,Xintai1}, and near-field propagation conditions \cite{Lipeng5}. The spatial agility afforded by MAs has been further advanced by the emergence of six-dimensional MAs (6DMAs), a generalized architecture that jointly harnesses translational displacement and rotational orientation to enhance wavefront manipulation \cite{Xiandan3,Xiandan4}. From a system-level optimization perspective, a substantial body of work has focused on the deliberate placement of antenna elements to improve user fairness and spectral utilization in both uplink \cite{Guojie1,Zhenyu1,Guojie2} and downlink \cite{Lipeng6,Yifei1,Songjie1,Ziyuan1} multiuser environments, as well as channel acquisition based on limited measurements \cite{Wenyan_CE1,Wenyan_CE2}. To cope with the inherent non-convexity of MA position design, researchers have turned to a variety of numerical strategies, ranging from gradient-driven updates and alternating optimization frameworks to learning-aided methods. Concurrently, efforts to combine MA technology with emerging concepts such as integrated sensing and communication (ISAC) \cite{Wenyan_TSP,Wanting,Wu_ISAC} and intelligent reflecting surfaces (IRS) \cite{Weixin,Haoze,Wu_MIS,MA_IRS,MA_IRS1,MA_IRS2,MA_IRS3,peng2025double,peng2025single} have gained momentum, further highlighting the adaptability of MAs across diverse wireless paradigms \cite{Lipeng_Survey}. In parallel, a variety of flexible antenna technologies, such as fluid antennas \cite{KKWong1}, pinching antennas \cite{DZG}, and rotatable antennas \cite{peng2025rotatable}, have been extensively investigated to improve the performance of wireless networks.

Despite the extensive body of work dedicated to MA systems, a critical practical consideration remains largely unaddressed in the literature, i.e., the temporal cost associated with antenna repositioning. The preponderance of existing studies operates under the implicit premise that the duration of a transmission block is sufficiently long to render the antenna movement delay negligible. While this may hold in quasi-static environments, it becomes a tenuous assumption in more dynamic scenarios where the service interval for users is constrained. Furthermore, the antenna movement process itself may interfere with normal data transmission due to practical factors such as electromagnetic radiation generated by channel fluctuation and Doppler effect \cite{Lipeng2}. Consequently, a dedicated antenna movement phase must be allocated, during which no information transmission occurs. In such contexts, the time expended to physically relocate an antenna from its initial position to a newly optimized location constitutes a non-negligible overhead, directly curtailing the interval available for data transmission \cite{Honghao}. This introduces an inherent system trade-off: Relocating an antenna to a more propitious spatial coordinate can augment the attainable channel quality, yet the associated transit time erodes the effective communication throughput. Consequently, an optimization strategy that focuses solely on maximizing a channel metric (e.g., signal-to-noise ratio or channel gain) without accounting for movement latency may inadvertently select a target position that is too remote, thereby yielding a non-negligible degradation in throughput.

While the study in \cite{Honghao} pioneered the consideration of movement delay, it differs fundamentally from our work in system configuration, optimization methodology, and derived insights. First, \cite{Honghao} focuses on user-side MA deployment under a per-coherence-block movement protocol, whereas we investigate BS-side MA deployment with a single movement phase per service interval, significantly reducing actuation frequency. Second, the algorithmic strategies diverge: \cite{Honghao} employs the successive convex approximation (SCA)-based iterative surrogate optimization, while we develop a penalty-based alternating framework and a low-complexity fitting approach that bypasses exhaustive duration search. Third, the theoretical contributions are distinct yet complementary: \cite{Honghao} establishes path-count-dependent movement conditions for receiver-side placement, whereas we derive a closed-form speed threshold that determines when remaining stationary is optimal at the BS side. These multifaceted distinctions motivate our problem formulation and solution design, elaborated below.

Specifically, the main contributions of this paper are summarized as follows:
\begin{itemize}
    \item \textbf{Problem Formulation}: We formulate a novel minimum effective throughput maximization problem for the MAs-aided multiuser downlink system, which jointly optimizes the antenna movement duration, antenna deployment, power allocation, and transmit beamforming under practical constraints including finite service time, maximum movement speed, and minimum antenna spacing. This formulation explicitly captures the inherent tradeoff between the time invested in repositioning antennas for channel enhancement and the remaining time available for data transmission.

    \item \textbf{Algorithm Design}: To tackle the resulting non-convex problem, we propose two solutions with distinct complexities. The general method performs the one-dimensional search over movement durations, wherein a penalty-based alternating optimization algorithm integrating projected gradient descent (PGD) and closed-form auxiliary updates is employed to design the antenna positions. To reduce computational overhead, we further devise a low-complexity fitting method that samples only a few rate-duration pairs, fits either a quadratic or sigmoidal growth model, and obtains optimized movement durations instantly. A specialized two-antenna two-user case study validates the fundamental tradeoff and demonstrates the accuracy of the proposed fitting models.

    \item \textbf{Theoretical Insight}: We derive a closed-form condition on the maximum antenna movement speed under the general setup, i.e., if the movement speed falls below a derived threshold, the optimal strategy is to keep the antennas stationary for the entire transmission period. This insight provides a simple yet powerful criterion for real-time decision-making on whether to trigger BS-side antenna repositioning. Also, the accuracy of the derived threshold is further validated through a two-antenna two-user case study.

    \item \textbf{Performance Validation}: Extensive numerical simulations demonstrate that our algorithms substantially outperform static deployments and fixed-movement benchmarks, approach the instantaneous-movement upper bound at the high antenna movement speed, and exhibit remarkable robustness to variations in region size and number of antennas. The fitting method achieves pretty good performance with significantly reduced complexity, confirming its practical viability.
\end{itemize}

The rest of this paper is organized as follows. Section II introduces the system model and formulates the joint optimization problem for maximizing the minimum effective throughput. Section III proposes two methods with different complexities to solve the formulated problem. Section IV investigates the special case to provide analytical insights and validate the fitting models. Section V derives a closed-form condition on the maximum movement speed under which keeping antennas stationary is optimal. Section VI presents simulation results to verify the effectiveness of the proposed schemes. Finally, conclusions are drawn in Section VII.

 \textit{Notations}: For a complex scalar $a$, $a^*$ and ${\mathop{\rm Im}\nolimits}(a)$ denote its complex conjugate and imaginary part, respectively. For a complex vector $\mathbf{a}$, $\mathbf{a}^H$, $\mathbf{a}^T$, and $\|\mathbf{a}\|$ denote its conjugate transpose, transpose, and Euclidean norm, respectively. For a matrix $\mathbf{A}$, $\mathbf{A}^H$, $\mathbf{A}^{-1}$, $\operatorname{tr}(\mathbf{A})$, and $[\mathbf{A}]_{i,j}$ denote its conjugate transpose, inverse, trace, and the element at the $i$-th row and $j$-th column, respectively. $\mathbf{I}_N$ is the $N \times N$ identity matrix. $\mathcal{CN}(0,\sigma^2)$ denotes a circularly symmetric complex Gaussian distribution with zero mean and variance $\sigma^2$. $\mathbb{E}[\cdot]$ is the expectation operator. $\nabla_{\mathbf{x}} f$ denotes the gradient of function $f$ with respect to (w.r.t.) vector $\mathbf{x}$. For a time-dependent vector $\mathbf{a}(t)$, $\dot{\mathbf{a}}(t)$ denotes its derivative w.r.t. time. $\mathbb{R}$ and $\mathbb{C}$ denote the sets of real and complex numbers, respectively.

 \newcounter{mytempeqncnt}
\section{System Model and Problem Formulation}
\subsection{System Model}
As illustrated in Fig. 1, we consider the MAs-enabled multiuser downlink communication system, where the BS equipped with $N$ MAs serves $K$ single FPA-enabled users during a finite time interval $T$. At the BS, these $N$ MAs are positioned within a two-dimensional plane ${\cal L} \buildrel \Delta \over = L \times L$, where ${{\bf{a}}_n} = {[{x_n},{y_n}]^T}$ denotes the position of the $n$-th MA relative to the reference point ${[0,0]^T}$, with $n \in {\cal N} \buildrel \Delta \over = \left\{ {1,...,N} \right\}$ and ${\bf{A}} \buildrel \Delta \over = \left\{ {{{\bf{a}}_1},{{\bf{a}}_2},...,{{\bf{a}}_N}} \right\} \in {{\mathbb{R}}^{2 \times N}}$.

To fully exploit the potential of MA technology, the BS should perform antenna deployment optimization at the beginning of the time interval to reconfigure wireless channel conditions. However, the antenna movement process may interfere with normal data transmission due to factors such as channel fluctuation and
Doppler effect. In light of these, the total interval should be divided into two phases:

i) \textbf{Antenna movement phase}: In this phase, the BS needs to carefully adjust antenna deployment to create better channel conditions for subsequent data transmission. The duration of this phase is denoted as $t_{\rm {mov}}$, with ${t_{{\rm{mov}}}} \in [0,T)$. Concurrently, the optimized position of the $n$-th antenna within this duration is denoted as ${\bf{a}}_n^{{t_{{\rm{mov}}}}}$, with ${{\bf{A}}^{{t_{{\rm{mov}}}}}} \buildrel \Delta \over = \left\{ {{\bf{a}}_1^{{t_{{\rm{mov}}}}},...,{\bf{a}}_N^{{t_{{\rm{mov}}}}}} \right\}$. Once antenna positions are optimized, they remain fixed in the subsequent data transmission phase.

ii) \textbf{Data transmission phase}: In this phase with the duration $T - {t_{{\rm{mov}}}}$, the BS performs data transmission to the $K$ users based on the optimized antenna deployment pattern ${{\bf{A}}^{{t_{{\rm{mov}}}}}}$.

Since the moving region of MAs (on the order of several wavelengths) is significantly smaller than the signal propagation distance, the far-field channel condition between the BS and users is assumed. Therefore, considering the line-of-sight (LoS) propagation environment, the channel vector between the BS and user $k$ with the given antenna deployment pattern ${{{\bf{A}}^{{t_{{\rm{move}}}}}}}$ can be expressed as
\begin{equation}
\begin{split}{}
{{\bf{h}}_k}({{\bf{A}}^{{t_{{\rm{mov}}}}}}) = \sqrt {{\beta _k}} \left[ {{e^{j\frac{{2\pi }}{\lambda }{{({\bf{a}}_1^{{t_{{\rm{mov}}}}})}^T}{{\bf{b}}_k}}},...,{e^{j\frac{{2\pi }}{\lambda }{{({\bf{a}}_N^{{t_{{\rm{mov}}}}})}^T}{{\bf{b}}_k}}}} \right],
\end{split}
\end{equation}
where ${{\beta _k}}$ is the large-scale fading coefficient, $\lambda $ is the carrier wavelength, ${{\bf{b}}_k} = {\left[ {\cos {\theta _k}\sin {\phi _k},\sin {\theta _k}} \right]^T}$, with ${\theta _k} \in [ - \pi /2,\pi /2]$ and ${\phi _k} \in [ - \pi /2,\pi /2]$ denoting the elevation and azimuth angles of departure (AoDs) for user $k$, respectively.

During the data transmission phase, let $P_k$ and ${{\bf{w}}_k} \in {{\mathbb{C}}^{N \times 1}}$ respectively denote the transmit power and the beamforming vector at the BS for serving user $k$, satisfying $\left\| {{{\bf{w}}_k}} \right\| = 1$, $\forall k \in {\cal K} \buildrel \Delta \over = \left\{ {1,...,K} \right\}$. Then, the received signal at user $k$ can be expressed as
\begin{equation}
\begin{split}{}
{y_k} = & \sqrt {{P_k}} {{\bf{h}}_k}({{\bf{A}}^{{t_{{\rm{mov}}}}}}){{\bf{w}}_k}{s_k}\\
 & + \sum\nolimits_{j = 1,j \ne k}^K {\sqrt {{P_j}} {{\bf{h}}_k}({{\bf{A}}^{{t_{{\rm{mov}}}}}}){{\bf{w}}_j}{s_j} + {n_k}},
\end{split}
\end{equation}
where $s_k$ is the information-bearing signal for user $k$ with ${\mathbb{E}}\left[ {{{\left| {{s_k}} \right|}^2}} \right] = 1$, $\forall k \in {\cal K}$, and ${n_k} \sim {\cal{CN}}(0,{\sigma ^2})$ is the additive white Gaussian noise (AWGN) at user $k$. Based on (2), the received signal-to-interference-plus-noise
ratio (SINR) of user $k$ is given by
\begin{equation}
\begin{split}{}
{\gamma _k}\left( {{{\bf{A}}^{{t_{{\rm{mov}}}}}},{\bf{P}},{\bf{W}}} \right) = \frac{{{P_k}{{\left| {{{\bf{h}}_k}({{\bf{a}}^{{t_{{\rm{mov}}}}}}){{\bf{w}}_k}} \right|}^2}}}{{\sum\nolimits_{j = 1,j \ne k}^K {{P_j}{{\left| {{{\bf{h}}_k}({{\bf{a}}^{{t_{{\rm{mov}}}}}}){{\bf{w}}_j}} \right|}^2} + {\sigma ^2}} }},
\end{split}
\end{equation}
where ${\bf{P}} \buildrel \Delta \over = \left\{ {{P_1},...,{P_K}} \right\}$ and ${\bf{W}} \buildrel \Delta \over = \left\{ {{{\bf{w}}_1},...,{{\bf{w}}_K}} \right\} \in {{\mathbb{C}}^{N \times K}}$. Therefore, the achievable rate of user $k$ (in bits/s/Hz) is
\begin{equation}
\begin{split}{}
{R_k}\left( {{{\bf{A}}^{{t_{{\rm{mov}}}}}},{\bf{P}},{\bf{W}}} \right) = {\log _2}\left( {1 + {\gamma _k}\left( {{{\bf{A}}^{{t_{{\rm{mov}}}}}},{\bf{P}},{\bf{W}}} \right)} \right).
\end{split}
\end{equation}

Based on (4) and considering the antenna movement delay, the effective throughput of user $k$ (in bits/Hz) over the time interval $T$ can be derived as
\begin{equation}
\begin{split}{}
{\rm{T}}{{\rm{h}}_k}\left( {{{\bf{A}}^{{t_{{\rm{mov}}}}}},{\bf{P}},{\bf{W}}} \right) = (T - {t_{{\rm{mov}}}}){R_k}\left( {{{\bf{A}}^{{t_{{\rm{mov}}}}}},{\bf{P}},{\bf{W}}} \right).
\end{split}
\end{equation}

\subsection{Problem Formulation}
In this paper, with a focus on fairness, we aim to maximize the minimum effective throughput among all users, by jointly optimizing the antenna movement duration $t_{\rm mov}$, the antenna deployment pattern ${{{\bf{A}}^{{t_{{\rm{mov}}}}}}}$, the power allocations ${\bf{P}}$ and the transmit beamforming vectors ${\bf{W}}$. The resulting problem can be formulated as
 \begin{align}
&({\rm{P1}}):{\rm{  }}\mathop {\max }\limits_{{t_{{\rm{mov}}}},{{\bf{A}}^{{t_{{\rm{mov}}}}}},{\bf{P}},{\bf{W}}} \ \mathop {\min }\limits_{k \in {\cal K}} \left\{ {{\rm{T}}{{\rm{h}}_k}\left( {{{\bf{A}}^{{t_{{\rm{mov}}}}}},{\bf{P}},{\bf{W}}} \right)} \right\} \tag{${\rm{6a}}$}\\
{\rm{              }}&{\rm{s.t.}} \ \ \sum\nolimits_{k = 1}^K {{P_k}}  \le {P_{{\rm{tot}}}},\tag{${\rm{6b}}$}\\
 &\ \ \ \ \ \left\| {{{\bf{w}}_k}} \right\| = 1,\forall k \in {\cal K},\tag{${\rm{6c}}$}\\
 &\ \ \ \ \ \left\| {{\bf{a}}_n^{{t_{{\rm{mov}}}}} - {\bf{a}}_n^{\rm {{Initial}}}} \right\| \le {V_{\max }}{t_{{\rm{mov}}}},\forall n \in {\cal N}, \tag{${\rm{6d}}$}\\
 &\ \ \ \ \ \left\| {{\bf{a}}_n^{{t_{{\rm{mov}}}}} - {\bf{a}}_m^{{t_{{\rm{mov}}}}}} \right\| \ge {d_{\min }},\forall n,m \in {\cal N}, n \ne m, \tag{${\rm{6e}}$}\\
  &\ \ \ \ \ \ {\bf{a}}_n^{{t_{{\rm{mov}}}}} \in {\cal L}, \forall n \in {\cal N}, \tag{${\rm{6f}}$}
\end{align}
where ${P_{{\rm{tot}}}}$ in (6b) denotes the total transmit power budget of the BS, ${{\bf{a}}_n^{{\rm{Initial}}}}$ and ${V_{\max }}$ in (6d) denote the initial position of the $n$-th antenna and the maximum antenna movement speed, respectively. More specifically, the constraints in (6d) indicate that the movement distance of any antenna within the interval $[0,{t_{{\rm{mov}}}}]$ should not exceed ${V_{\max }}{t_{{\rm{mov}}}}$. The constraints in (6e) impose a minimum distance ${d_{\min }}$ between any two distinct antennas at the BS to avoid mutual coupling effects.

\textbf{Remark 1:} From problem (P1), we can observe a fundamental tradeoff in determining the antenna movement duration: i) On one hand, a larger $t_{\rm{mov}}$ allows each antenna to move in a wider region, thereby increasing the likelihood of reaching a more favorable position and creating better channel conditions for subsequent data transmission; ii) On the other hand, as $t_{\rm{mov}}$ increases, the remaining time available for data transmission inevitably decreases, which is evidently detrimental to improving the effective throughput. Consequently, optimizing $t_{\rm{mov}}$ plays a crucial role in this context. However, due to the strong coupling among all variables in the objective of (P1) and the interdependence between the antenna deployment pattern ${{\bf{A}}^{{t_{{\rm{mov}}}}}}$ and $t_{\rm{mov}}$ through the constraints in (6d), developing an efficient algorithm to directly solve (P1) is challenging.

\section{Solutions to (P1)}
To simplify the above highly non-convex problem, we adopt the zero-forcing (ZF)-based transmit beamforming at the BS. This choice is motivated by two primary reasons. First, the ZF-based beamforming offers a low implementation complexity, particularly for the large $N$ \cite{Haiquan}, and can also achieve favorable fairness when combined with adaptive power allocations. Second, based on the ZF-based beamforming, the objective can be simplified into a form that depends only on $t_{\rm{mov}}$ and ${{\bf{A}}^{{t_{{\rm{mov}}}}}}$, thereby facilitating analytical insights.\footnote{Alternatively, the BS may adopt other forms of transmit beamforming, such as weighted minimum mean square error (WMMSE). In that case, problem (P1) remains solvable, albeit through more complex procedures. Since the primary focus of this paper is to reveal the fundamental tradeoff in determining the antenna movement duration rather than optimizing beamforming, we leave the investigation of more complex scenarios to future work.}

Specifically, given ${{{\bf{A}}^{{t_{{\rm{mov}}}}}}}$, the ZF-based transmit beamforming for user $k$ can be expressed as ${\bf{w}}_k^{{\rm{ZF}}}({{\bf{A}}^{{t_{{\rm{mov}}}}}}) = {{\bf{w}}_k}({{\bf{A}}^{{t_{{\rm{mov}}}}}})/\left\| {{{\bf{w}}_k}({{\bf{A}}^{{t_{{\rm{mov}}}}}})} \right\|$ \cite{Haiquan}, with
\begin{equation}
\setcounter{equation}{7}
\begin{split}{}
{{\bf{w}}_k}({{\bf{A}}^{{t_{{\rm{mov}}}}}}) = \left( {{{\bf{I}}_N} - \overline {{{\bf{B}}_k}} ({{\bf{A}}^{{t_{{\rm{mov}}}}}})} \right){\bf{h}}_k^H({{\bf{A}}^{{t_{{\rm{mov}}}}}}),
\end{split}
\end{equation}
where
\begin{equation} \nonumber
\begin{split}{}
\overline {{{\bf{B}}_k}} ({{\bf{A}}^{{t_{{\rm{mov}}}}}}) =& {{\bf{B}}_k}({{\bf{A}}^{{t_{{\rm{mov}}}}}})\\
& \times {\left( {{\bf{B}}_k^H({{\bf{A}}^{{t_{{\rm{mov}}}}}}){{\bf{B}}_k}({{\bf{A}}^{{t_{{\rm{mov}}}}}})} \right)^{ - 1}}{\bf{B}}_k^H({{\bf{A}}^{{t_{{\rm{mov}}}}}}),
\end{split}
\end{equation}
and ${{\bf{B}}_k}({{\bf{A}}^{{t_{{\rm{mov}}}}}}) = {\left[ {\begin{array}{*{20}{l}}
{{{\bf{h}}_1}({{\bf{A}}^{{t_{{\rm{mov}}}}}});...;{{\bf{h}}_{k - 1}}({{\bf{A}}^{{t_{{\rm{mov}}}}}});}\\
{{{\bf{h}}_{k + 1}}({{\bf{A}}^{{t_{{\rm{mov}}}}}});...;{{\bf{h}}_K}({{\bf{A}}^{{t_{{\rm{mov}}}}}})}
\end{array}} \right]^H} \in {{\mathbb{C}}^{N \times (K - 1)}}$.

Substituting ${\left\{ {{\bf{w}}_k^{{\rm{ZF}}}({{\bf{A}}^{{t_{{\rm{mov}}}}}})} \right\}_{k \in {\cal K}}}$ into (3), the SINR of user $k$ can be simplified as \cite{Haiquan}
\begin{equation}
\begin{split}{}
{\gamma _k}\left( {{{\bf{A}}^{{t_{{\rm{mov}}}}}},{\bf{P}}} \right) = \frac{{{P_k}}}{{{{\left[ {{{\left( {{{\bf{H}}^H}({{\bf{A}}^{{t_{{\rm{mov}}}}}}){\bf{H}}({{\bf{A}}^{{t_{{\rm{mov}}}}}})} \right)}^{ - 1}}} \right]}_{k,k}}{\sigma ^2}}},
\end{split}
\end{equation}
with ${\bf{H}}({{\bf{A}}^{{t_{{\rm{mov}}}}}}) = {\left[ {{{\bf{h}}_1}({{\bf{A}}^{{t_{{\rm{mov}}}}}});...;{{\bf{h}}_K}({{\bf{A}}^{{t_{{\rm{mov}}}}}})} \right]^H} \in {{\mathbb C}^{N \times K}}$.

Furthermore, given ${{{\bf{A}}^{{t_{{\rm{mov}}}}}}}$, to ensure fairness, the optimal power allocations at the BS obviously should be the solutions of the following equations:
\begin{equation}
\begin{split}{}
\left\{ {\begin{array}{*{20}{c}}
{{\gamma _1}\left( {{{\bf{A}}^{{t_{{\rm{mov}}}}}},{\bf{P}}} \right) = ... = {\gamma _K}\left( {{{\bf{A}}^{{t_{{\rm{mov}}}}}},{\bf{P}}} \right)}\\
{\sum\nolimits_{k = 1}^K {{P_k}}  = {P_{{\rm{tot}}}}}
\end{array}} \right.,
\end{split}
\end{equation}
which yields the optimal power for user $k$ as
\begin{equation}
\begin{split}{}
P_k^* = \frac{{{{\left[ {{{\left( {{{\bf{H}}^H}({{\bf{A}}^{{t_{{\rm{mov}}}}}}){\bf{H}}({{\bf{A}}^{{t_{{\rm{mov}}}}}})} \right)}^{ - 1}}} \right]}_{k,k}}}}{{\sum\nolimits_{j = 1}^K {{{\left[ {{{\left( {{{\bf{H}}^H}({{\bf{A}}^{{t_{{\rm{mov}}}}}}){\bf{H}}({{\bf{A}}^{{t_{{\rm{mov}}}}}})} \right)}^{ - 1}}} \right]}_{j,j}}} }}{\mkern 1mu} {P_{{\rm{tot}}}}, \forall k.
\end{split}
\end{equation}

Substituting $P_k^*$ into (8), the SINR of user $k$ for all $\forall k \in {\cal K}$ can be derived as
\begin{equation}
\begin{split}{}
&{\gamma _k}\left( {{{\bf{A}}^{{t_{{\rm{mov}}}}}}} \right) = \frac{{{P_{{\rm{tot}}}}}}{{\sum\nolimits_{j = 1}^K {{{\left[ {{{\left( {{{\bf{H}}^H}({{\bf{A}}^{{t_{{\rm{mov}}}}}}){\bf{H}}({{\bf{A}}^{{t_{{\rm{mov}}}}}})} \right)}^{ - 1}}} \right]}_{j,j}}} {\sigma ^2}}}\\
 =& \frac{{{P_{{\rm{tot}}}}}}{{{\rm{tr}}\left( {{{\left( {{{\bf{H}}^H}({{\bf{A}}^{{t_{{\rm{mov}}}}}}){\bf{H}}({{\bf{A}}^{{t_{{\rm{mov}}}}}})} \right)}^{ - 1}}} \right){\sigma ^2}}} \buildrel \Delta \over = \gamma \left( {{{\bf{A}}^{{t_{{\rm{mov}}}}}}} \right), \forall k.
\end{split}
\end{equation}

Based on (11), problem (P1) can be simplified into a formulation that involves only ${{t_{{\rm{mov}}}}}$ and ${{{\bf{A}}^{{t_{{\rm{mov}}}}}}}$, i.e.,
 \begin{align}
&({\rm{P2}}):{\rm{  }}\mathop {\max }\limits_{{t_{{\rm{mov}}}},{{\bf{A}}^{{t_{{\rm{mov}}}}}}} \ (T - {t_{{\rm{mov}}}}){\log _2}\left( {1 + \gamma \left( {{{\bf{A}}^{{t_{{\rm{mov}}}}}}} \right)} \right) \tag{${\rm{12a}}$}\\
{\rm{              }}&{\rm{s.t.}} \ \ \ \ \left\| {{\bf{a}}_n^{{t_{{\rm{mov}}}}} - {\bf{a}}_n^{\rm {{Initial}}}} \right\| \le {V_{\max }}{t_{{\rm{mov}}}},\forall n \in {\cal N}, \tag{${\rm{12b}}$}\\
 &\ \ \ \ \ \ \ \ \left\| {{\bf{a}}_n^{{t_{{\rm{mov}}}}} - {\bf{a}}_m^{{t_{{\rm{mov}}}}}} \right\| \ge {d_{\min }},\forall n,m \in {\cal N}, n \ne m, \tag{${\rm{12c}}$}\\
  &\ \ \ \ \ \ \ \ \ {\bf{a}}_n^{{t_{{\rm{mov}}}}} \in {\cal L}, \forall n \in {\cal N}. \tag{${\rm{12d}}$}
\end{align}

Despite this simplification, the main difficulty in solving (P2) still lies in the coupling between the movement duration ${{t_{{\rm{mov}}}}}$ and the antenna deployment ${{{\bf{A}}^{{t_{{\rm{mov}}}}}}}$, which appears in both the objective and the constraints in (12b). In the following, we will propose two methods with different complexity levels to effectively solve (P2).

\begin{figure*}[b!]
  \hrulefill
\setcounter{equation}{15}
\begin{equation}
\begin{split}{}
{\bf{a}}_n^{{t_{{\rm{mov}}}},(i)} = {{\cal P}_{{{\cal C}_n}}}\left\{ {{\bf{a}}_n^{{t_{{\rm{mov}}}},(i - 1)} - {\eta ^{(i - 1)}}{\nabla _{{\bf{a}}_n^{{t_{{\rm{mov}}}}}}}f({{\bf{A}}^{{t_{{\rm{mov}}}},(i - 1)}}) - 2\rho ({\bf{a}}_n^{{t_{{\rm{mov}}}},(i - 1)} - {{\bf{z}}_n})} \right\}.
\end{split}
\end{equation}
\end{figure*}

\subsection{The General Method for Solving (P2)}
In this subsection, we propose a general method to solve (P2). The core idea is to perform the one-dimensional search over ${t_{{\rm{mov}}}} \in [0,T)$. Subsequently, for each given $t_{\rm{mov}}$, problem (P2) should be solved to obtain the desirable antenna deployment.

Specifically, with the given ${t_{{\rm{mov}}}}$, by ignoring irrelevant items and noting that: i) maximizing ${\log _2}\left( {1 + \gamma \left( {{{\bf{A}}^{{t_{{\rm{mov}}}}}}} \right)} \right)$ is equivalent to maximizing ${\gamma \left( {{{\bf{A}}^{{t_{{\rm{mov}}}}}}} \right)}$; ii) $\gamma \left( {{{\bf{A}}^{{t_{{\rm{mov}}}}}}} \right)$ is monotonically decreasing w.r.t. ${{\rm{tr}}\left( {{{\left( {{{\bf{H}}^H}({{\bf{A}}^{{t_{{\rm{mov}}}}}}){\bf{H}}({{\bf{A}}^{{t_{{\rm{mov}}}}}})} \right)}^{ - 1}}} \right)}$, (P2) can be simplified as
 \begin{align}
&({\rm{P2.1}}):{\rm{  }}\mathop {\min }\limits_{{{\bf{A}}^{{t_{{\rm{mov}}}}}}} \ {{\rm{tr}}\left( {{{\left( {{{\bf{H}}^H}({{\bf{A}}^{{t_{{\rm{mov}}}}}}){\bf{H}}({{\bf{A}}^{{t_{{\rm{mov}}}}}})} \right)}^{ - 1}}} \right)} \tag{${\rm{13a}}$}\\
{\rm{              }}&{\rm{s.t.}} \ \ \ \ \left\| {{\bf{a}}_n^{{t_{{\rm{mov}}}}} - {\bf{a}}_n^{\rm {{Initial}}}} \right\| \le {V_{\max }}{t_{{\rm{mov}}}},\forall n \in {\cal N}, \tag{${\rm{13b}}$}\\
 &\ \ \ \ \ \ \ \ \left\| {{\bf{a}}_n^{{t_{{\rm{mov}}}}} - {\bf{a}}_m^{{t_{{\rm{mov}}}}}} \right\| \ge {d_{\min }},\forall n,m \in {\cal N}, n \ne m, \tag{${\rm{13c}}$}\\
  &\ \ \ \ \ \ \ \ \ {\bf{a}}_n^{{t_{{\rm{mov}}}}} \in {\cal L}, \forall n \in {\cal N}. \tag{${\rm{13d}}$}
\end{align}

To handle the non-convex distance constraints in (13c), we adopt a penalty-based framework \cite{Penalty}. Specifically, we introduce auxiliary variables $\mathbf{z}_n$ corresponding to each MA position $\mathbf{a}_n^{t_{\mathrm{mov}}}$ and reformulate (P2.1) as the following penalized problem:
\begin{align}
&{\rm{(P2.2)}}:\mathop {\min }\limits_{\mathbf{A}^{t_{\mathrm{mov}}},\{\mathbf{z}_n\}} \ f({{\bf{A}}^{{t_{{\rm{mov}}}}}}) + \rho {\sum\nolimits_{n = 1}^N {\left\| {{\bf{a}}_n^{{t_{{\rm{mov}}}}} - {{\bf{z}}_n}} \right\|} ^2} \tag{14a}\\
{\rm{              }}&{\rm{s.t.}} \ \ \ \ \left\| {{\bf{a}}_n^{{t_{{\rm{mov}}}}} - {\bf{a}}_n^{\rm {{Initial}}}} \right\| \le {V_{\max }}{t_{{\rm{mov}}}},\forall n \in {\cal N}, \tag{${\rm{14b}}$}\\
  &\ \ \ \ \ \ \ \ \ {\bf{a}}_n^{{t_{{\rm{mov}}}}} \in {\cal L}, \forall n \in {\cal N}, \tag{${\rm{14c}}$}\\
&\ \  \ \quad \quad \|\mathbf{z}_n - \mathbf{z}_m\| \geq d_{\min},\quad \forall n,m \in \mathcal{N}, n \neq m, \tag{14d}
\end{align}
where $\rho>0$ is a penalty factor and $f({{\bf{A}}^{{t_{{\rm{mov}}}}}}) \buildrel \Delta \over = {\rm{tr}}({({{\bf{H}}^H}({{\bf{A}}^{{t_{{\rm{mov}}}}}}){\bf{H}}({{\bf{A}}^{{t_{{\rm{mov}}}}}}))^{ - 1}})$. In this reformulation, the original non-convex distance constraints are imposed only on the auxiliary variables $\{\mathbf{z}_n\}$, while the original variables $\left\{ {{\bf{a}}_n^{{t_{{\rm{mov}}}}}} \right\}$ retain only the convex constraints (14b) and (14c), i.e., the mobility limit and the box constraint $\mathcal{L}$. Problem (P2.2) is then solved by alternating optimization (AO), with $\rho$ gradually increased in each outer iteration to enforce $\mathbf{a}_n^{t_{\mathrm{mov}}}\approx\mathbf{z}_n$. The details of AO are presented in the follows.

\textbf{Optimization of $\mathbf{A}^{t_{\mathrm{mov}}}$}: For fixed $\{\mathbf{z}_n\}$, the subproblem reduces to
\begin{align}
&\mathop {\min }\limits_{\mathbf{A}^{t_{\mathrm{mov}}}} \ f({{\bf{A}}^{{t_{{\rm{mov}}}}}}) + \rho {\sum\nolimits_{n = 1}^N {\left\| {{\bf{a}}_n^{{t_{{\rm{mov}}}}} - {{\bf{z}}_n}} \right\|} ^2} \tag{15a}\\
{\rm{              }}&{\rm{s.t.}} \ \ \ \ \left\| {{\bf{a}}_n^{{t_{{\rm{mov}}}}} - {\bf{a}}_n^{\rm {{Initial}}}} \right\| \le {V_{\max }}{t_{{\rm{mov}}}},\forall n \in {\cal N}, \tag{${\rm{15b}}$}\\
  &\ \ \ \ \ \ \ \ \ {\bf{a}}_n^{{t_{{\rm{mov}}}}} \in {\cal L}, \forall n \in {\cal N}. \tag{${\rm{15c}}$}
\end{align}
This smooth optimization over a convex feasible set can be efficiently solved using the PGD method. Specifically, let $\eta^{(i-1)}$ denote the step size (learning rate) at the $i-1$-th inner iteration, which can be chosen as a constant (e.g., $10^{-3}$) or a diminishing sequence. Then the update for the position of antenna $n$ at the $i$-th inner iteration is given in (16), where ${\nabla_{{\bf{a}}_n^{{t_{{\rm{mov}}}}}}}f({{\bf{A}}^{{t_{{\rm{mov}}}},(i - 1)}})$ denotes the gradient of $f({{\bf{A}}^{{t_{{\rm{mov}}}},(i - 1)}})$ w.r.t. $\mathbf{a}_n^{t_{\mathrm{mov}}}$, the expression of which is derived in Appendix A. In addition, ${{\cal P}_{{\cal C}_n}}$ in (16) denotes the Euclidean projection onto the convex set ${\cal C}_n = {\cal L} \cap {\rm{O}}({\bf{a}}_n^{{\rm{Initial}}},{V_{\max }}{t_{{\rm{mov}}}})$, where ${\rm{O}}({\bf{a}}_n^{{\rm{Initial}}},{V_{\max }}{t_{{\rm{mov}}}})$ is the closed disk centered at ${\bf{a}}_n^{{\rm{Initial}}}$ with radius ${V_{\max }}{t_{{\rm{mov}}}}$.

For a given ${\bf{p}} = {[{p_x},{p_y}]^T}$, its projection onto ${\cal C}_n$, i.e., ${{\cal P}_{{\cal C}_n}}\left\{ {\bf{p}} \right\}$, can be obtained by solving the following problem:
\begin{align}
&\mathop {\min }\limits_{{\bf{x}}} {\left\| {{\bf{x}} - {\bf{p}}} \right\|^2} \tag{17a}\\
&{\rm{s.t.}} \ {\bf{x}} \in {\cal L}, \tag{17b}\\
  & \quad \ {\left\| {{\bf{x}} - {\bf{a}}_n^{{\rm{Initial}}}} \right\|^2} \le {\left( {{V_{\max }}{t_{{\rm{mov}}}}} \right)^2}, \tag{17c}
\end{align}
which is convex and can be directly solved via the standard convex optimization tools such as CVX.

\textbf{Optimization of $\{\mathbf{z}_n\}$}: For fixed $\mathbf{A}^{t_{\mathrm{mov}}}$, the subproblem becomes
\begin{align}
&\mathop {\min }\limits_{\{ {{\bf{z}}_n}\} } {\sum\nolimits_{n = 1}^N {\left\| {{{\bf{z}}_n} - {\bf{a}}_n^{{t_{{\rm{mov}}}}}} \right\|} ^2} \tag{18a}\\
&{\rm{s.t.}} \quad \|\mathbf{z}_n - \mathbf{z}_m\| \geq d_{\min},\quad \forall n,m \in \mathcal{N}, n \neq m. \tag{18b}
\end{align}
This subproblem is identical to $\mathcal{P}1$-b in \cite{Penalty} and can be solved optimally by the closed-form procedure described therein (Algorithm 2). Since the derivation is fully presented in \cite{Penalty}, we omit the details here for brevity.

In summary, the detailed procedures for solving (P2.1) with the given ${t_{{\rm{mov}}}}$ are concluded in Algorithm 1, where we denote the optimized position of the $n$-th antenna as ${{\bf{a}}_{n,{\rm{opt}}}^{{t_{{\rm{mov}}}}}}$, with ${\bf{A}}_{{\rm{opt}}}^{{t_{{\rm{mov}}}}} \buildrel \Delta \over = \left\{ {{\bf{a}}_{1,{\rm{opt}}}^{{t_{{\rm{mov}}}}},...,{\bf{a}}_{N,{\rm{opt}}}^{{t_{{\rm{mov}}}}}} \right\}$.

\begin{algorithm}
\caption{Penalty-Based Optimization Framework for Solving (P2.1)}
\begin{algorithmic}[1]
\State \textbf{Initialization:} Set initial antenna positions $\left\{ {{\bf{a}}_n^{{\rm{Initial}}}} \right\}$, auxiliary variables $\{\mathbf{z}_n\}$ and penalty factor $\rho>0$.
\State \textbf{Repeat:}
\State \quad \textbf{Step 1 (Optimize $\mathbf{A}^{t_{\mathrm{mov}}}$):} For fixed $\{\mathbf{z}_n\}$, iteratively update each antenna position using the PGD according to (16) until convergence.
\State \quad \textbf{Step 2 (Optimize $\{\mathbf{z}_n\}$):} For fixed $\mathbf{A}^{t_{\mathrm{mov}}}$, solve the distance-constrained subproblem (18) optimally via the closed-form procedure given in \cite{Penalty}.
\State \quad \textbf{Step 3 (Update $\rho$):} Increase the penalty factor: $\rho \leftarrow \rho \times c$ with $c>1$.
\State \textbf{Until:} Stopping criterion is satisfied, e.g., $\max_n \|\mathbf{a}_n^{t_{\mathrm{mov}}} - \mathbf{z}_n\| \leq \epsilon$ or maximum outer iterations reached.
\State \textbf{Output:} Return the optimized antenna positions ${{\bf{A}}_{{\rm{opt}}}^{{t_{{\rm{mov}}}}}}$.
\end{algorithmic}
\end{algorithm}

 With a given $t_{\mathrm{mov}}$ and optimized antenna deployment ${\bf{A}}_{{\rm{opt}}}^{{t_{{\rm{mov}}}}}$, defining $\chi ({t_{{\rm{mov}}}},{\bf{A}}_{{\rm{opt}}}^{{t_{{\rm{mov}}}}})$ as the corresponding objective value of (P2). In light of the fundamental tradeoff discussed in Remark 1, $\chi ({t_{{\rm{mov}}}},{\bf{A}}_{{\rm{opt}}}^{{t_{{\rm{mov}}}}})$ is expected to increase initially and then decrease with $t_{\mathrm{mov}}$. Therefore, the optimized $t_{\mathrm{mov}}$ and ${{\bf{A}}_{{\rm{opt}}}^{{t_{{\rm{mov}}}}}}$ can be identified by evaluating $\chi ({t_{{\rm{mov}}}},{\bf{A}}_{{\rm{opt}}}^{{t_{{\rm{mov}}}}})$ over the one-dimensional grid of $t_{\mathrm{mov}}$.

\textbf{Complexity Analysis}: In the inner iteration (Step 1) of Algorithm 1, computing the gradient $\nabla_{\mathbf{a}_n^{t_{\mathrm{mov}}}}f(\mathbf{A}^{t_{\mathrm{mov}}})$ (derived in Appendix A) requires forming $\mathbf{H}^H(\mathbf{A}^{t_{\mathrm{mov}}})\mathbf{H}(\mathbf{A}^{t_{\mathrm{mov}}})$ with the complexity of $\mathcal{O}(NK^2)$, inverting the resulting $K\times K$ matrix with the complexity of $\mathcal{O}(K^3)$, and then applying the chain rule to combine $\partial\bigl((\mathbf{H}^H(\mathbf{A}^{t_{\mathrm{mov}}})\mathbf{H}(\mathbf{A}^{t_{\mathrm{mov}}}))^{-1}\bigr)/\partial \mathbf{a}_n^{t_{\mathrm{mov}}}$ with the derivative $\partial\mathbf{H}(\mathbf{A}^{t_{\mathrm{mov}}})/\partial\mathbf{a}_n^{t_{\mathrm{mov}}}$. Since $\mathbf{H}(\mathbf{A}^{t_{\mathrm{mov}}})$ depends only on $\mathbf{a}_n^{t_{\mathrm{mov}}}$ when computing the gradient w.r.t. the $n$-th antenna, the overall complexity per antenna is $\mathcal{O}(NK^2+K^3)$. Consequently, the gradients for all $N$ antennas are obtained with the complexity of $\mathcal{O}(N^2K^2+NK^3)$ per PGD iteration. The projection $\mathcal{P}_{\mathcal{C}}$ onto the convex set $\mathcal{C}$ is implemented using convex optimization tools and requires $\mathcal{O}(N)$ operations. Concurrently, the auxiliary variables $\{\mathbf{z}_n\}$ are updated in closed form (Step 2) following the method in \cite{Penalty}, with the complexity of $\mathcal{O}(N^2)$. Let $I_{\mathrm{pgd}}$ and $I_{\mathrm{ao}}$ denote the numbers of inner PGD iterations and outer AO iterations, respectively. Then the total complexity for a fixed $t_{\mathrm{mov}}$ is
$\mathcal{O}\Bigl(I_{\mathrm{ao}}\bigl(I_{\mathrm{pgd}}(N^2K^2+NK^3)+N^2\bigr)\Bigr)$. By further denoting $\delta$ as the accuracy of searching ${{t_{{\rm{mov}}}}}$, the total complexity of solving (P2) is about ${\cal O}({I_{{\rm{ao}}}}({I_{{\rm{pgd}}}}({N^2}{K^2} + N{K^3}) + {N^2})T/\delta )$.

\begin{algorithm}
\caption{The Fitting Method for Solving (P2)}
\begin{algorithmic}[1]
\State Solve (P2.1) without the speed constraint in (13b), and output $\left\{ {{\bf{a}}_n^*} \right\}_{n = 1}^N$. Then, based on $\left\{ {{\bf{a}}_n^*} \right\}_{n = 1}^N$ and $t_l$ in (17), determine ${t_{{\rm{mov,max}}}}$.

\State Uniformly samples $S$ rate-movement duration pairs in the interval $[0,{t_{{\rm{mov,max}}}}]$.

\State Based on the sampled data, use the nlinfit function in MATLAB to generate the fitting model $g({t_{{\rm{mov}}}})$ that approximates ${\log _2}\left( {1 + \gamma \left( {{\bf{A}}_{{\rm{opt}}}^{{t_{{\rm{mov}}}}}} \right)} \right)$ for ${t_{{\rm{mov}}}} \in [0,{t_{{\rm{mov,max}}}}]$.

\State Use the simple one-dimensional search to optimize $(T - {t_{{\rm{mov}}}})g({t_{{\rm{mov}}}})$ over ${t_{{\rm{mov}}}} \in [0,{t_{{\rm{mov,max}}}}]$, thereby obtaining the optimized ${t_{{\rm{mov}}}}$.
\end{algorithmic}
\end{algorithm}


\begin{figure*}
    \centering
    \begin{minipage}{0.48\linewidth}
        \centering
        \includegraphics[width=7.5cm]{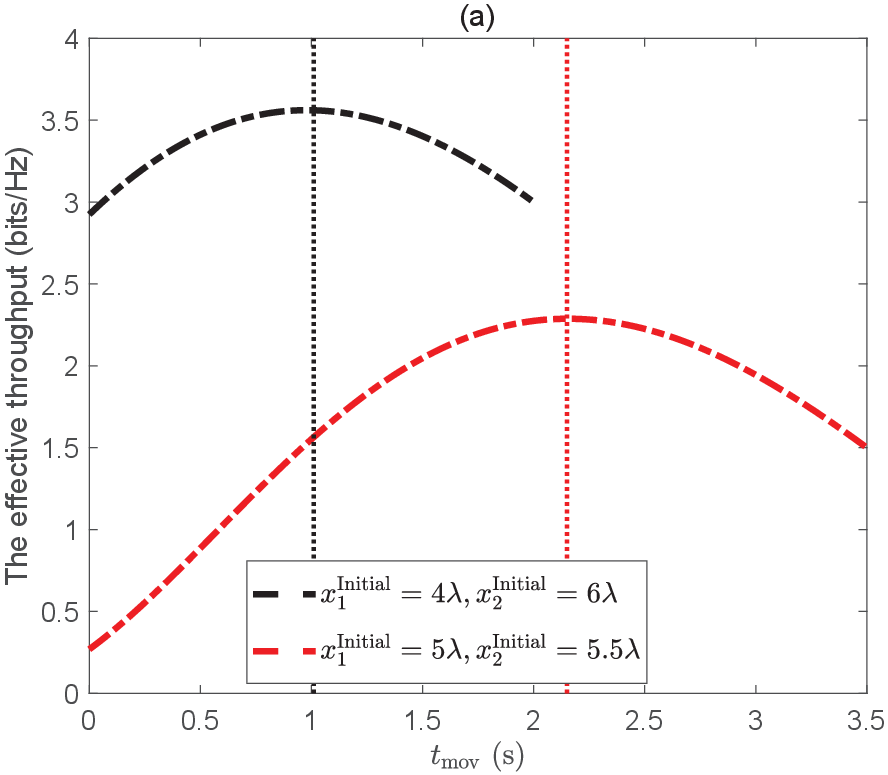}
    \end{minipage}
    \hfill
    \begin{minipage}{0.48\linewidth}
        \centering
        \includegraphics[width=7.5cm]{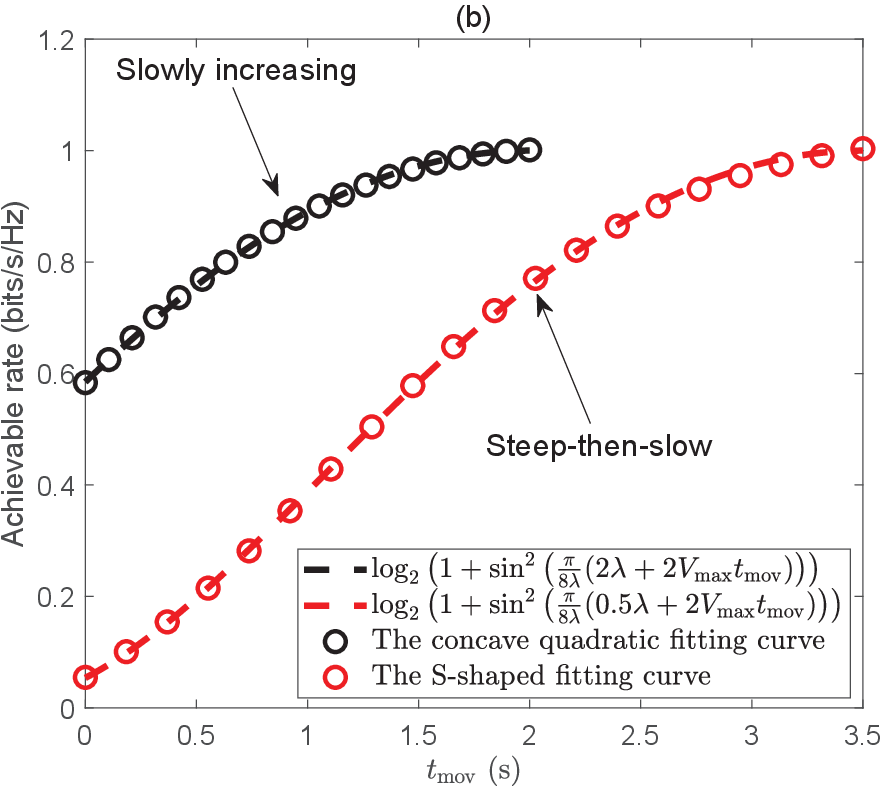}
    \end{minipage}
    \captionsetup{font=small}
    \caption{(a) Effective throughput versus $t_{\rm{mov}}$ for Case i) (black dashed) and Case ii) (red dashed), showing the fundamental tradeoff; (b) Achievable rate versus $t_{\rm{mov}}$: Exact curves (dashed) and their quadratic (black circle) and sigmoidal (red circle) fits with $S = 5$ samples.}
    \label{fig:2}
       \vspace{-10pt}
\end{figure*}

\begin{figure*}[b!]
  \hrulefill
\setcounter{equation}{22}
\begin{equation}
\begin{split}{}
&{\rm{tr}}\left( {{{\left( {{{\bf{H}}^H}({{\bf{x}}^{{t_{{\rm{mov}}}}}}){\bf{H}}({{\bf{x}}^{{t_{{\rm{mov}}}}}})} \right)}^{ - 1}}} \right) = {\rm{tr}}\left( {{{\left( {\left[ {\begin{array}{*{20}{c}}
{{{\bf{h}}_1}({{\bf{x}}^{{t_{{\rm{mov}}}}}})}\\
{{{\bf{h}}_2}({{\bf{x}}^{{t_{{\rm{mov}}}}}})}
\end{array}} \right]\left[ {\begin{array}{*{20}{c}}
{{\bf{h}}_1^H({{\bf{x}}^{{t_{{\rm{mov}}}}}})}&{{\bf{h}}_2^H({{\bf{x}}^{{t_{{\rm{mov}}}}}})}
\end{array}} \right]} \right)}^{ - 1}}} \right)\\
 =& {\rm{ tr}}\left( {{{\left( {\beta \left[ {\begin{array}{*{20}{c}}
2&{{e^{j\varepsilon x_1^{{t_{\rm{mov}}}}}} + {e^{j\varepsilon x_2^{{t_{\rm{mov}}}}}}}\\
{{e^{ - j\varepsilon x_1^{{t_{\rm{mov}}}}}} + {e^{ - j\varepsilon x_2^{{t_{\rm{mov}}}}}}}&2
\end{array}} \right]} \right)}^{ - 1}}} \right) \mathop  = \limits^{(a)} \frac{1}{{\beta {{\sin }^2}\left( {\frac{\varepsilon }{2}(x_1^{{t_{{\rm{mov}}}}} - x_2^{{t_{{\rm{mov}}}}})} \right)}}.
\end{split}
\end{equation}
\end{figure*}

\subsection{The Low-Complexity Fitting Method for Solving (P2)}
The general method described in the previous subsection yields a high-quality solution, yet it incurs substantial computational complexity, primarily due to the exhaustive search over $t_{\mathrm{mov}}$. Furthermore, such complexity may become prohibitive in scenarios where real-time adaptation is required. To address this limitation, we propose a low-complexity fitting method that circumvents the need for exhaustive search.

To proceed, a crucial judgement is that the behavior of $\log_2\bigl(1 + \gamma(\mathbf{A}_{\mathrm{opt}}^{t_{\mathrm{mov}}})\bigr)$ as $t_{\mathrm{mov}}$ increases in $[0, T)$ can be categorized into two cases:
\begin{itemize}
    \item[i)] It continues to increase w.r.t. $t_{\mathrm{mov}} \in [0, T)$;
    \item[ii)] It increases and then saturates at $t_{\mathrm{mov,max}}$, with $t_{\mathrm{mov,max}} < T$.
\end{itemize}
In case ii), the effective throughput $(T - t_{\mathrm{mov}})\log_2\bigl(1 + \gamma(\mathbf{A}_{\mathrm{opt}}^{t_{\mathrm{mov}}})\bigr)$ will decrease monotonically for $t_{\mathrm{mov}} > t_{\mathrm{mov,max}}$ due to the reduction in data transmission time, making further consideration unnecessary. Therefore, the search for the optimized $t_{\mathrm{mov}}$ should be restricted to the interval $[0, t_{\mathrm{mov,max}}]$.

  A key problem in the above is how to determine whether case i) or case ii) occurs. To this end, we first solve problem (P2.1) without the speed constraint in (13b), i.e., we allow unlimited movement speed to find the optimized antenna positions that minimize the trace objective. Denote the resulting deployment by $\{\mathbf{a}_n^*\}_{n=1}^N$. Then, the minimum time required for each antenna to travel from its initial position $\mathbf{a}_n^{\mathrm{Initial}}$ to $\mathbf{a}_n^*$ is given by $\|\mathbf{a}_n^* - \mathbf{a}_n^{\mathrm{Initial}}\|/V_{\mathrm{max}}$. Taking the maximum over all antennas, we obtain
\begin{equation}
\setcounter{equation}{19}
t_{l} = \frac{\max_n \|\mathbf{a}_n^* - \mathbf{a}_n^{\mathrm{Initial}}\|}{V_{\mathrm{max}}}.
\end{equation}

If $t_l \ge T$, it means that even if the entire transmission period $T$ is used for antenna movement, the antennas cannot reach the optimized positions $\{\mathbf{a}_n^*\}$ defined in the unconstrained case. Consequently, over the interval $t_{\mathrm{mov}}\in[0,T)$, the achievable rate $\log_2(1+\gamma(\mathbf{A}_{\mathrm{opt}}^{t_{\mathrm{mov}}}))$ keeps increasing with $t_{\mathrm{mov}}$ because a longer movement duration allows the antennas to get closer to the optimized deployment, and no saturation occurs within $T$. Hence case i) holds, and we set $t_{\mathrm{mov},\mathrm{max}}\triangleq T$. Otherwise, if $t_l < T$, the antennas are able to reach the optimized positions before the end of the transmission period. After that, further increasing $t_{\mathrm{mov}}$ does not improve the channel condition (the rate saturates). Therefore, case ii) is established, and the saturation point is exactly $t_l$. We thus set $t_{\mathrm{mov},\mathrm{max}}\triangleq t_l$.

After determining $t_{\mathrm{mov,max}}$ in cases i) and ii), we next formally build a low-complexity fitting model. Specifically, we define $g(t_{\mathrm{mov}})$ as an approximation of $\log_2(1+\gamma(\mathbf{A}_{\mathrm{opt}}^{t_{\mathrm{mov}}}))$, and aim to obtain a closed-form expression for $g(t_{\mathrm{mov}})$ w.r.t. ${t_{{\rm{mov}}}} \in [0,{t_{{\rm{mov,max}}}}]$. To proceed, based on the effective distance between the initial antenna positions and the optimized deployment $\{\mathbf{a}_n^*\}_{n=1}^N$, we discuss two typical growth patterns of $\log_2(1+\gamma(\mathbf{A}_{\mathrm{opt}}^{t_{\mathrm{mov}}}))$ w.r.t. $t_{\mathrm{mov}}$ within the interval $[0, t_{\mathrm{mov,max}}]$.
\begin{enumerate}
    \item[i)] \textbf{Slowly increasing}: This scenario occurs when the effective distance mentioned above is small. In this case, the antennas require only a short movement duration to reach their desired positions, and the achievable rate exhibits a gradually diminishing growth rate, i.e., its second derivative remains negative and decreases in magnitude. Accordingly, a concave quadratic function is adopted to characterize $g(t_{\mathrm{mov}})$ over $t_{\mathrm{mov}} \in [0, t_{\mathrm{mov,max}}]$:
    \begin{equation}
    g(t_{\mathrm{mov}}) = C_1 (t_{\mathrm{mov}} - C_2)^2 + C_3,
    \end{equation}
    where $C_1 < 0$, $C_2 > 0$, and $C_3 > 0$ are fitting parameters. The negative quadratic term ensures a concave shape with a decreasing growth rate. The parameter $C_2$ represents the movement duration at which the quadratic function attains its maximum value. In this case, $C_2 \ge t_{\mathrm{mov,max}}$, since $\log_2(1+\gamma(\mathbf{A}_{\mathrm{opt}}^{t_{\mathrm{mov}}}))$ is non-decreasing over $[0, t_{\mathrm{mov,max}}]$, meaning the rate either increases throughout the interval (if $C_2 > t_{\mathrm{mov,max}}$) or just reaches saturation at the endpoint (if $C_2 = t_{\mathrm{mov,max}}$). $C_3$ determines the maximum approximate rate at $t_{\mathrm{mov}} = C_2$, and $C_1$ controls the curvature.

     \item[ii)] \textbf{Steep-then-slow}: This scenario occurs when the effective distance is relatively large. Consequently, the antennas require a longer movement duration to approach the favorable configuration. The achievable rate first increases rapidly as antennas move toward the optimized region, and then gradually saturates as they converge. This corresponds to a second derivative that first increases and then decreases. An S-shaped (sigmoidal) function is employed to capture this growth pattern:
    \begin{equation}
    g(t_{\mathrm{mov}}) = C_1 + \frac{C_2}{1 + e^{-(C_3 + C_4 t_{\mathrm{mov}})}},
    \end{equation}
    where $C_1 > 0$, $C_2 > 0$, $C_3 < 0$, and $C_4 > 0$ are fitting parameters. Here, $C_1 + C_2$ is theoretical maximum rate (as $t_{\mathrm{mov}} \to +\infty$). Due to the finite movement duration, the actual achievable rate when ${t_{{\rm{mov}}}} = {t_{{\rm{mov,max}}}}$ satisfies ${C_1} + \frac{{{C_2}}}{{1 + {e^{ - ({C_3} + {C_4}{t_{{\rm{mov,max}}}})}}}} \le {\log _2}(1 + \gamma \left( {\left\{ {{\bf{a}}_n^*} \right\}} \right))$. $C_4 > 0$ controls the steepness of the transition. The inflection point is located at $t_{\mathrm{mov}} = -C_3/C_4$, therefore $C_3 < 0$ shifts the inflection point to the positive $t_{\mathrm{mov}}$ region, reflecting that a certain movement duration is needed before the rapid growth occurs.
\end{enumerate}

Once the fitting function is determined, the approximate objective $(T - t_{\mathrm{mov}})g(t_{\mathrm{mov}})$ can be optimized straightforwardly over $t_{\mathrm{mov}} \in [0, t_{\mathrm{mov,max}}]$ to obtain an optimized solution with significantly reduced complexity.

In practice, the fitting parameters can be obtained by uniformly sampling a small number of rate-movement duration pairs within the interval ${t_{{\rm{mov}}}} \in [0,{t_{{\rm{mov,max}}}}]$ and then applying nonlinear least squares regression, e.g., via the nlinfit function in MATLAB, to fit the chosen model (quadratic or sigmoidal) to the sampled data. This approach eliminates the need for exhaustive optimization across all possible $t_{\mathrm{mov}}$. The detailed procedures for solving (P2) based on the fitting method are presented in Algorithm 2.

\textbf{Complexity Analysis}: The computational complexity of the proposed fitting method primarily arises from two tasks: i) Solving the problem (P2.1) without constraints in (13b) once to obtain the optimized deployment $\{\mathbf{a}_n^*\}$, and ii) Solving the same problem for each of the $S$ sampled movement durations in the interval $[0,t_{\mathrm{mov,max}}]$. The complexity of each such optimization is identical to that of solving (P2.1) for a fixed $t_{\mathrm{mov}}$ in Algorithm 1. In contrast, the nonlinear least-squares fitting (e.g., using nlinfit) and the subsequent one-dimensional search over the closed-form expression $(T-t_{\mathrm{mov}})g(t_{\mathrm{mov}})$ incur negligible overhead and can be safely ignored. Therefore, the overall complexity of Algorithm 2 is approximately ${\cal O}\left( {(1 + S){I_{{\rm{ao}}}}({I_{{\rm{pgd}}}}({N^2}{K^2} + N{K^3}) + {N^2})} \right)$. Since the number of samples $S$ is typically much smaller than the number of grid points required by exhaustive search, Algorithm 2 achieves substantial complexity reduction, making it well suited for real-time applications.

\section{Insights from the Special Case of $N = K = 2$}
In this section, we consider a special case of $N = K = 2$ to better illustrate the aforementioned fundamental tradeoff while simultaneously verifying the effectiveness of the proposed fitting method. For simplification, these two antennas at the BS are assumed to move along the one-dimensional segment of length $L$.

Denote $x_n$ ($n = 1,2$) as the position of the $n$-th MA at the BS relative to the reference point 0, and ${\theta _k}$ ($k = 1,2$) the AoD for user $k$. Then the channel vector between the BS and user $k$ with the antenna vector ${{\bf{x}}^{{t_{{\rm{mov}}}}}} \buildrel \Delta \over = {[x_1^{{t_{\rm{mov}}}},x_2^{{t_{\rm{mov}}}}]^T}$ can be expressed as
\begin{equation}
\setcounter{equation}{22}
{{\bf{h}}_k}({{\bf{x}}^{{t_{{\rm{mov}}}}}}) = \sqrt {{\beta _k}} \left[ {{e^{j\frac{{2\pi }}{\lambda }x_1^{{t_{\rm{mov}}}}\cos {\theta _k}}},\;{e^{j\frac{{2\pi }}{\lambda }x_2^{{t_{\rm{mov}}}}\cos {\theta _k}}}} \right].
\end{equation}

Based on (22), the objective of (P2.1) can be first simplified as in (23), where $\varepsilon  = \frac{{2\pi }}{\lambda }(\cos {\theta _2} - \cos {\theta _1})$ and the equation (a) is established since for a Hermitian matrix $\left[ {\begin{array}{*{20}{c}}
a&b\\
{{b^H}}&a
\end{array}} \right]$, its inverse is $\frac{1}{{{a^2} - {{\left| b \right|}^2}}}\left[ {\begin{array}{*{20}{c}}
a&{ - b}\\
{ - {b^H}}&a
\end{array}} \right]$.

Based on (23), problem (P2) in this special case can be simplified as
\begin{align}
&({\rm{P3}}):{\rm{  }}\mathop {\max }\limits_{{t_{{\rm{mov}}}},{\left\{ {x_n^{{t_{\rm{mov}}}}} \right\}_{n = 1}^2}} \ (T - {t_{{\rm{mov}}}}) \notag \\
& \quad \quad \quad \quad \ \qquad \times {\log _2}\left( {1 + \gamma \left( {\left\{ {x_n^{{t_{\rm{mov}}}}} \right\}_{n = 1}^2} \right)} \right) \tag{24a} \\
&{\rm{s.t.}} \ \ \ \ \left| {x_n^{{t_{{\rm{mov}}}}} - x_n^{{\rm{Initial}}}} \right| \le {V_{\max }}{t_{{\rm{mov}}}},\forall n \in {\cal N}, \tag{24b}\\[-0.5\jot]
&\ \ \ \ \ \ \ \ \left| {x_1^{{t_{{\rm{mov}}}}} - x_2^{{t_{{\rm{mov}}}}}} \right| \ge {d_{\min }}, \tag{24c}\\[-0.5\jot]
&\ \ \ \ \ \ \ \ \ {{x}}_n^{{t_{{\rm{mov}}}}} \in [0,L], \forall n \in {\cal N}, \tag{24d}
\end{align}
where $\gamma \left( {\left\{ {x_n^{{t_{{\rm{mov}}}}}} \right\}_{n = 1}^2} \right) = \frac{{{P_{{\rm{tot}}}}\beta }}{{{\sigma ^2}}}{\sin ^2}\left( {\frac{\varepsilon }{2}(x_1^{{t_{{\rm{mov}}}}} - x_2^{{t_{\rm{mov}}}})} \right)$.

To clearly observe the fundamental tradeoff and validate the fitting method, we set the system parameters as $L = 10\lambda$, $V_{\max}=0.5\lambda$/s, $T=5$~s, $P_{\text{tot}}\beta/\sigma^2 = 1$, and $\varepsilon = \pi/(4\lambda)$. With these settings, the SINR in (P3) simplifies to
\begin{equation}
\setcounter{equation}{25}
\gamma\!\left(\{x_n^{t_{\rm{mov}}}\}_{n=1}^2\right) = \sin^2\!\left(\frac{\pi}{8\lambda}\bigl(x_1^{t_{\rm{mov}}} - x_2^{t_{\rm{mov}}}\bigr)\right),
\end{equation}
which achieves its global maximum when $\bigl|x_1^{t_{\rm{mov}}} - x_2^{t_{\rm{mov}}}\bigr| = 4\lambda$. To illustrate how the initial antenna spacing affects the growth pattern of the achievable rate and to validate the two proposed fitting models (quadratic and sigmoidal), we consider the following two representative cases.

\textbf{Case i) (Slowly increasing pattern):}  Set $x_1^{\rm{Initial}} = 4\lambda$, $x_2^{\rm{Initial}} = 6\lambda$, giving an initial spacing of $2\lambda$, which is smaller than the optimal $4\lambda$. Therefore, the two antennas must move in opposite directions to increase the spacing. The distance between them evolves as $d_{t_{\rm{mov}}} = 2\lambda + 2V_{\max}t_{\rm{mov}}$. Solving $d_{t_{\rm{mov}}} = 4\lambda$ yields $t_{\rm{mov},\max} = 2$~s. Substituting $d_{t_{\rm{mov}}}$ into the rate expression, problem (P3) reduces to (P3.1), i.e.,
\begin{align}
& ({\rm{P3.1}}):\ \mathop {\max }\limits_{{t_{{\rm{mov}}}}} \; (5 - t_{\rm{mov}}) \notag \\
& \qquad \times \log_2\left(1 + \sin^2\left(\frac{\pi}{8\lambda}(2\lambda + 2V_{\max}t_{\rm{mov}})\right)\right) \tag{26a} \\
& \text{s.t. } t_{\rm{mov}} \in [0,2]. \tag{26b}
\end{align}
In Fig.~2(a), the black dashed line shows the objective of (P3.1) as a function of ${t_{\rm{mov}}} \in [0,2]$, which first increases and then decreases, clearly verifying the fundamental tradeoff mentioned above. Fig.~2(b) plots the rate itself (black dashed line) together with the concave quadratic fit (black circle) obtained via (20) with $S = 5$ sampled pairs. The fitting parameters are ${C_1} =  - {\rm{0}}{\rm{.0975}}$, ${C_2} = {\rm{2}}{\rm{.0714}}$ and ${C_3} = {\rm{1}}{\rm{.0017}}$. The close match demonstrates the effectiveness of the quadratic fitting model for the slowly increasing regime.

\textbf{Case ii) (Steep-then-slow pattern):}
Now set $x_1^{\rm{Initial}} = 5\lambda$, $x_2^{\rm{Initial}} = 5.5\lambda$, i.e., an initial spacing of only $0.5\lambda$, which is also smaller than $4\lambda$. Hence, the two antennas must again move in opposite directions to enlarge the spacing. The distance evolves as $d_{t_{\rm{mov}}} = 0.5\lambda + 2V_{\max}t_{\rm{mov}}$. Solving $d_{t_{\rm{mov}}} = 4\lambda$ gives $t_{\rm{mov},\max} = 3.5$~s. The corresponding simplified problem reduces to (P3.2), i.e.,
\begin{align}
& ({\rm{P3.2}}):\ \mathop {\max }\limits_{{t_{{\rm{mov}}}}} \; (5 - t_{\rm{mov}}) \notag \\
& \qquad \times \log_2\left(1 + \sin^2\left(\frac{\pi}{8\lambda}(0.5\lambda + 2V_{\max}t_{\rm{mov}})\right)\right) \tag{27a} \\
& \text{s.t. } t_{\rm{mov}} \in [0,3.5]. \tag{27b}
\end{align}
In Fig.~2(a), the red dashed line depicts the objective of (P3.2) over ${t_{\rm{mov}}} \in [0,3.5]$, which also exhibits a clear tradeoff. Fig.~2(b) shows the rate itself (red dashed line) and the sigmoidal fit (red circle) obtained via (21) with $S = 5$ sampled pairs. The fitting parameters are ${C_1} =  - {\rm{0}}{\rm{.1465}}$, ${C_2} = {\rm{1}}{\rm{.1959}}$, ${C_3} = - {\rm{1}}{\rm{.5977}}$ and ${C_4} = {\rm{1}}{\rm{.3763}}$. The red circle closely matches the red dashed line, confirming the accuracy of the sigmoidal fitting model for the steep-then-slow growth pattern.

Together, these two cases, i.e., one with a slowly increasing rate and the other with a steep-then-slow increase, demonstrate the necessity and accuracy of both proposed fitting models (quadratic and sigmoidal) under different growth behaviors, thereby validating the comprehensiveness of our low-complexity fitting approach.

\section{In What Conditions Moving Antennas is Adverse?}
The preceding sections have revealed a fundamental tradeoff in MAs-aided systems: A longer movement duration ${{t_{{\rm{mov}}}}}$ allows antennas to reach more favorable positions, yet it reduces the time left for data transmission. However, an even more fundamental question remains unexplored: Is antenna movement always beneficial? In other words, under what conditions does the potential gain in channel quality fail to compensate for the loss of transmission time, making movement actually harmful? Answering this question is of both theoretical and practical importance. It provides a simple criterion for the system to quickly decide whether to trigger antenna repositioning or simply keep the antennas static.

Motivated by this, we in this section aim to derive a closed-form condition on the maximum antenna movement speed threshold ${V_{\rm{th}}}$, such that the optimal strategy is to remain stationary when ${V_{\max }} \le {V_{\rm{th}}}$. The details are presented in Proposition 1.

\textbf{Proposition 1 (Stationary Antenna Optimality):} Define the SINR of each user with the initial antenna positions ${{\bf{A}}^{\rm{Initial}}} \buildrel \Delta \over = \left\{ {{\bf{a}}_1^{\rm{Initial}},{\bf{a}}_2^{\rm{Initial}},...,{\bf{a}}_N^{\rm{Initial}}} \right\}$ as $\gamma ({{\bf{A}}^{\rm{Initial}}})$, the expression of which can be obtained based on (11), and the resulted rate $R({{\bf{A}}^{\rm{Initial}}}) = {\log _2}(1 + \gamma ({{\bf{A}}^{\rm{Initial}}}))$. Further, let
\begin{equation}
\setcounter{equation}{28}
{V_{\rm{th}}} = \frac{{R({{\bf{A}}^{\rm{Initial}}})}}{{T\sum\nolimits_{n = 1}^N {\left\| {{\nabla _{{{\bf{a}}_n}}}R({{\bf{A}}^{\rm{Initial}}})} \right\|} }},
\end{equation}
where $\nabla_{\mathbf{a}_n}R(\mathbf{A}^{\mathrm{Initial}})$ denotes the gradient of $R$ evaluated at the initial positions. Then, if the maximum antenna movement speed satisfies ${V_{\max }} \le {V_{\rm{th}}}$, the optimal solution of (P2) is $t_{\rm{mov}}^* = 0$, i.e., the antennas remain stationary throughout the whole transmission period, and the maximum effective throughput is $TR({{\bf{A}}^{\rm{Initial}}})$. Otherwise, it is beneficial to allocate a certain duration for antenna movement, and the optimized movement duration and antenna deployment can then be obtained using the methods presented in Sections III and IV.
\begin{proof}
Recall the objective of problem (P2), where the effective throughput for a given movement duration $t_{\mathrm{mov}}$ and the optimized deployed antenna positions $\mathbf{A}_{\mathrm{opt}}^{t_{\mathrm{mov}}}$ is defined as
\begin{equation}
\chi (t_{\mathrm{mov}},\mathbf{A}_{\mathrm{opt}}^{t_{\mathrm{mov}}}) = (T - t_{\mathrm{mov}})\log_2\!\left(1 + \gamma\bigl(\mathbf{A}_{\mathrm{opt}}^{t_{\mathrm{mov}}}\bigr)\right).
\end{equation}
For brevity, denote $\chi (t_{\mathrm{mov}}) \triangleq \chi (t_{\mathrm{mov}},\mathbf{A}_{\mathrm{opt}}^{t_{\mathrm{mov}}})$. Clearly, $\chi (0) = T\log_2(1+\gamma(\mathbf{A}^{\mathrm{Initial}}))$ and $\chi (T)=0$. As explained by the fundamental tradeoff in Remark~1 and numerically verified in Fig.~2(a), $\chi (t_{\mathrm{mov}})$ is unimodal on $[0,T]$: It either increases first and then decreases, or is monotonically decreasing. This shape arises because the rate ${\log _2}\left( {1 + \gamma \left( {{\bf{A}}_{{\rm{opt}}}^{{t_{{\rm{mov}}}}}} \right)} \right)$ is bounded above while the remaining time $T-t_{\mathrm{mov}}$ shrinks linearly.

Therefore, to prove that $t_{\mathrm{mov}}=0$ is the global maximizer, it suffices to show that the right derivative of $\chi (t_{\mathrm{mov}})$ at $t_{\mathrm{mov}}=0$ is non-positive. If ${\left. {\frac{{\partial \chi }}{{\partial {t_{{\rm{mov}}}}}}} \right|_{{t_{{\rm{mov}}}} = {0^ + }}} \le 0$, the unimodal property forces $\chi (t_{\mathrm{mov}})$ to be non-increasing over the whole interval, hence $\chi (t_{\mathrm{mov}}) \le \chi (0)$ for all $t_{\mathrm{mov}}\in[0,T]$.

\textbf{Derivative at $t_{\mathrm{mov}}=0^+$:}
For small $t_{\mathrm{mov}}$, the optimized antenna positions $\mathbf{A}_{\mathrm{opt}}^{t_{\mathrm{mov}}}$ can be approximated by a smooth trajectory $\mathbf{a}_n(t_{\mathrm{mov}})$ with initial velocity $\mathbf{v}_n = \dot{\mathbf{a}}_n(0^+)$ satisfying $\|\mathbf{v}_n\|\le V_{\max}$, $\forall n \in {\cal N}$. Differentiating $\chi (t_{\mathrm{mov}})$ with $t_{\mathrm{mov}}$ gives
\begin{equation}
\begin{split}{}
&{\left. {\frac{{\partial \chi }}{{\partial {t_{{\rm{mov}}}}}}} \right|_{{t_{{\rm{mov}}}} = 0}} \\
=& - R({{\bf{A}}^{{\rm{Initial}}}}) + T\sum\nolimits_{n = 1}^N {{{\left( {{\nabla _{{{\bf{a}}_n}}}R({{\bf{A}}^{\rm{Initial}}})} \right)}^T}{{\bf{v}}_n}}.
\end{split}
\end{equation}
Because $\chi (t_{\mathrm{mov}})$ is the maximum effective throughput achievable with movement duration $t_{\mathrm{mov}}$, the initial velocities $\{\mathbf{v}_n\}$ are chosen to maximize the right-hand side (the system can move antennas in any admissible direction). Hence,
\begin{equation}
\begin{split}{}
&{\left. {\frac{{\partial \chi }}{{\partial {t_{{\rm{mov}}}}}}} \right|_{{t_{{\rm{mov}}}} = {0^ + }}} \\
=& \mathop {\max }\limits_{\left\| {{{\bf{v}}_n}} \right\| \le {V_{\max }}} \left\{ \begin{array}{l}
 - R({{\bf{A}}^{{\rm{Initial}}}})\\
 + T\sum\nolimits_{n = 1}^N {{{\left( {{\nabla _{{{\bf{a}}_n}}}R({{\bf{A}}^{{\rm{Initial}}}})} \right)}^T}{{\bf{v}}_n}}
\end{array} \right\}.
\end{split}
\end{equation}
 The maximum is attained by aligning each $\mathbf{v}_n$ with its gradient direction:
\begin{equation}
\mathbf{v}_n = V_{\max}\,\frac{\nabla_{\mathbf{a}_n}R(\mathbf{A}^{\mathrm{Initial}})}{\|\nabla_{\mathbf{a}_n}R(\mathbf{A}^{\mathrm{Initial}})\|}.
\end{equation}
Substituting this choice into (31) yields
\begin{equation}
\begin{split}{}
{\left. {\frac{{\partial \chi }}{{\partial {t_{{\rm{mov}}}}}}} \right|_{{t_{{\rm{mov}}}} = {0^ + }}} =& -R(\mathbf{A}^{\mathrm{Initial}}) \\
&+ T V_{\max} \sum_{n=1}^{N} \|\nabla_{\mathbf{a}_n}R(\mathbf{A}^{\mathrm{Initial}})\|.
\end{split}
\end{equation}

To have ${\left. {\frac{{\partial \chi }}{{\partial {t_{{\rm{mov}}}}}}} \right|_{{t_{{\rm{mov}}}} = {0^ + }}} \le 0$, we require
\begin{equation}
-R(\mathbf{A}^{\mathrm{Initial}}) + T V_{\max} \sum_{n=1}^{N} \|\nabla_{\mathbf{a}_n}R(\mathbf{A}^{\mathrm{Initial}})\| \le 0,
\end{equation}
which is equivalent to
\begin{equation}
V_{\max} \le \frac{R(\mathbf{A}^{\mathrm{Initial}})}{T \sum_{n=1}^{N} \|\nabla_{\mathbf{a}_n}R(\mathbf{A}^{\mathrm{Initial}})\|} \triangleq V_{\mathrm{th}}.
\end{equation}

Therefore, if $V_{\max} \le V_{\mathrm{th}}$, then ${\left. {\frac{{\partial \chi }}{{\partial {t_{{\rm{mov}}}}}}} \right|_{{t_{{\rm{mov}}}} = {0^ + }}}\le 0$, and by the unimodality of $\chi (t_{\mathrm{mov}})$ we conclude that $t_{\mathrm{mov}}=0$ is optimal. This completes the proof.
\end{proof}

\textbf{Remark 2:} Proposition 1 provides a threshold for the maximum antenna movement speed. Similarly, a threshold for the total time interval can be derived. Specifically, by enforcing ${\left. {\frac{{\partial \chi }}{{\partial {t_{{\rm{mov}}}}}}} \right|_{{t_{{\rm{mov}}}} = {0^ + }}} \le 0$ and solving for $T$, we obtain
\begin{equation}
{T_{\rm{th}}} = \frac{{R({{\bf{A}}^{\rm{Initial}}})}}{{{V_{\max }}\sum\nolimits_{n = 1}^N {\left\| {{\nabla _{{{\bf{a}}_n}}}R({{\bf{A}}^{\rm{Initial}}})} \right\|} }}.
\end{equation}
That is to say, if the actual transmission period satisfies $T \le {T_{\rm{th}}}$, then the right derivative of the effective throughput at $t_{\mathrm{mov}}=0$ is non-positive. Together with the unimodal nature of $\chi ({t_{{\rm{mov}}}})$, this also implies that $t_{\rm{mov}}^* = 0$ is optimal when $T \le {T_{\rm{th}}}$. In other words, when the total transmission time is sufficiently short, the time cost of moving antennas outweighs the potential rate gain, making it preferable to keep the antennas stationary.

 \begin{figure}
\centering
\includegraphics[width=7.5cm]{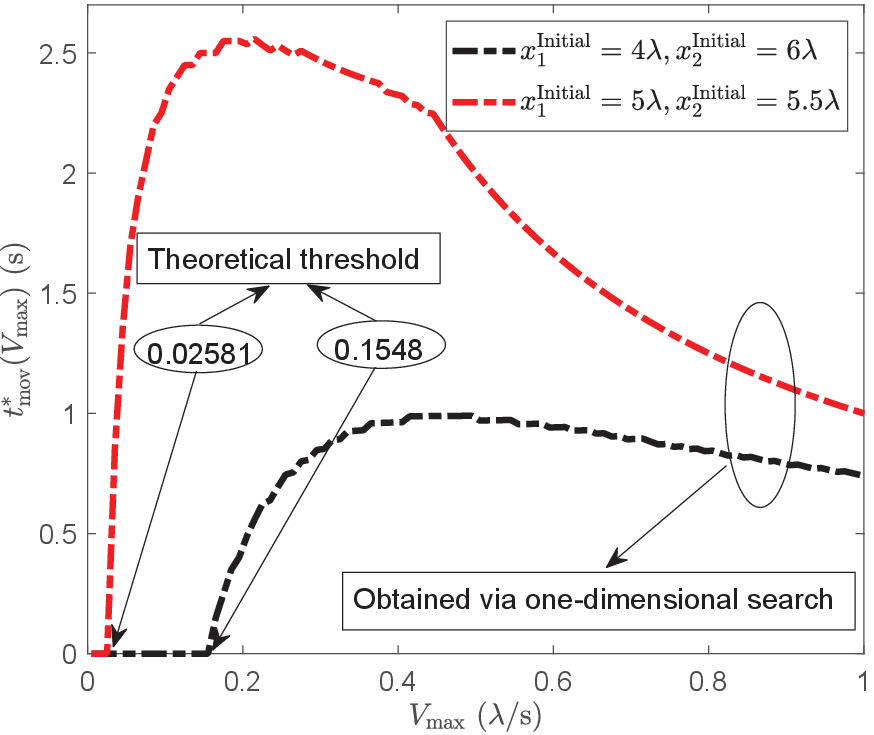}
\captionsetup{font=small}
\caption{The optimal movement duration $t_{{\rm{mov}}}^*({V_{\max }})$ w.r.t. ${V_{\max }}$ in the special case of $N = K = 2$.} \label{fig:Fig1}
\vspace{-10pt}
\end{figure}

\begin{figure*}
    \centering
    \begin{minipage}{0.48\linewidth}
        \centering
        \includegraphics[width=7.6cm]{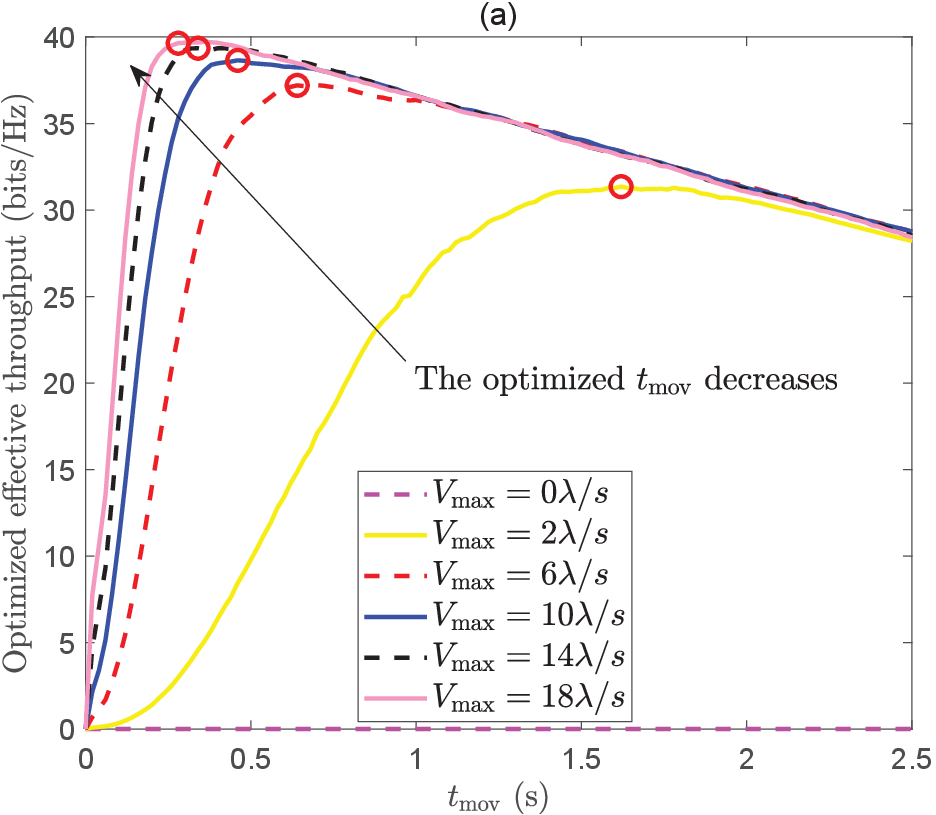}
    \end{minipage}
    \hfill
    \begin{minipage}{0.48\linewidth}
        \centering
        \includegraphics[width=7.6cm]{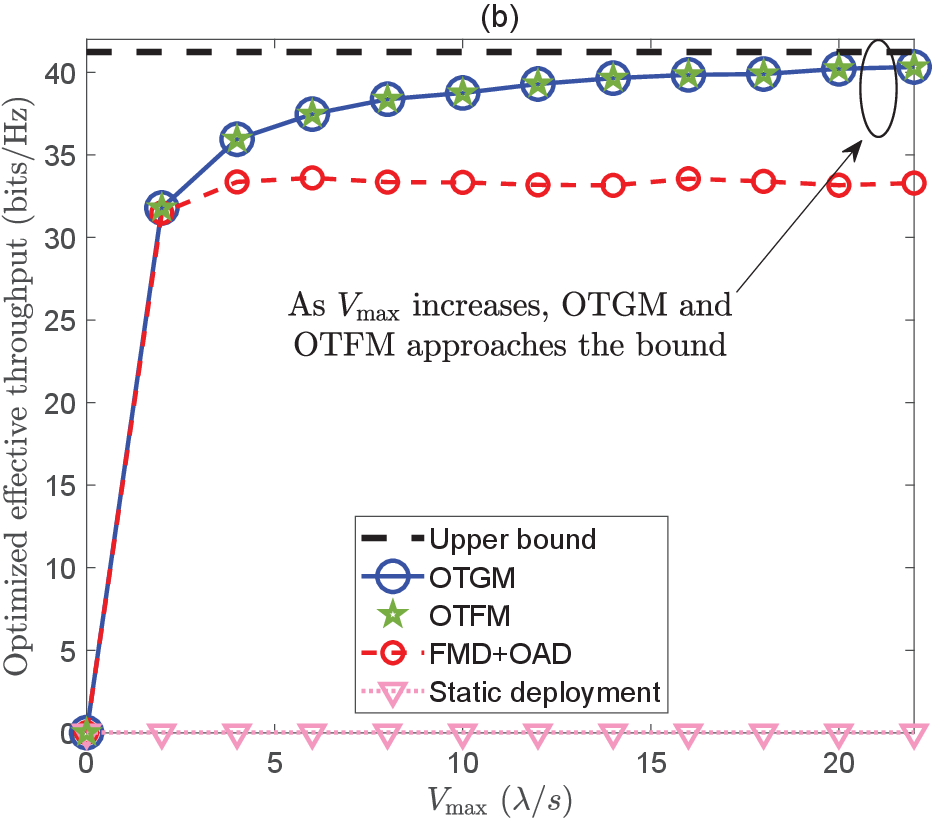}
    \end{minipage}
    \captionsetup{font=small}
    \caption{(a) Optimized effective throughput of OTGM w.r.t. ${t_{\rm{mov}}}$ for different ${V_{\max }}$; (b) Optimized effective throughput w.r.t ${V_{\max }}$ under different schemes.}
    \label{fig:2}
       \vspace{-10pt}
\end{figure*}

\textbf{Verification of Proposition 1:} To validate Proposition~1, we consider the special case $N = K = 2$ studied in Section~IV. Two different initial antenna configurations are examined. For each configuration, we first compute the corresponding theoretical threshold $V_{\mathrm{th}}$ using (28) and then perform numerical simulations to confirm that the stationary strategy is optimal when $V_{\max}\le V_{\mathrm{th}}$ and that movement becomes beneficial when $V_{\max}>V_{\mathrm{th}}$.

\textbf{Case~i): $x_1^{\mathrm{Initial}}=4\lambda,\; x_2^{\mathrm{Initial}}=6\lambda$}. In this case the initial spacing is $|\Delta_0|=2\lambda$ with $\Delta_0 = x_1^{\mathrm{Initial}}-x_2^{\mathrm{Initial}} = -2\lambda$. Further, let $\Delta = x_1^{t_{\mathrm{mov}}}-x_2^{t_{\mathrm{mov}}}$. The derivative of $R(\{ x_n^{{t_{{\rm{mov}}}}}\} _{n = 1}^2) = {\log _2}\left( {1 + \gamma (\{ x_n^{{t_{{\rm{mov}}}}}\} _{n = 1}^2)} \right)$ w.r.t. $\Delta$ based on (25) can be derived as
\begin{equation}
\frac{dR}{d\Delta}= \frac{1}{\ln2}\frac{1}{1+\sin^2(\pi\Delta/(8\lambda))}\frac{\pi}{8\lambda}\sin\!\left(\frac{\pi\Delta}{4\lambda}\right).
\end{equation}

Then, by substituting ${\Delta _0}$ into (37), we can derive
\begin{equation}
\left.\frac{dR}{d\Delta}\right|_{\Delta_0}= -\frac{1}{\ln2}\frac{1}{1+0.5}\frac{\pi}{8\lambda}= -\frac{\pi}{12\lambda\ln2}.
\end{equation}

The gradient w.r.t. each antenna is
\begin{equation}
\nabla_{x_1}R = \frac{dR}{d\Delta},\qquad \nabla_{x_2}R = -\frac{dR}{d\Delta}.
\end{equation}
Therefore
\begin{equation}
\sum\nolimits_{n = 1}^2 {\left| {{\nabla _{{x_n}}}R\left( {\left\{ {x_n^{{\rm{Initial}}}} \right\}_{n = 1}^2} \right)} \right|}  = 2\left| {\frac{{dR}}{{d\Delta }}} \right| = \frac{\pi }{{6\lambda \ln 2}}.
\end{equation}

The initial rate is
\begin{equation}
\begin{split}{}
R\left( {\left\{ {x_n^{{\rm{Initial}}}} \right\}_{n = 1}^2} \right) =& {\log _2}\left( {1 + {{\sin }^2}\frac{\pi }{4}} \right)\\
 =& {\log _2}(1.5)\;{\rm{bits/s/Hz}}.
\end{split}
\end{equation}
With $T=5$~s, (28) yields
\begin{equation}
\begin{split}{}
{V_{{\rm{th}}}} =& \frac{{R\left( {\left\{ {x_n^{{\rm{Initial}}}} \right\}_{n = 1}^2} \right)}}{{T\sum\nolimits_{n = 1}^2 {\left| {{\nabla _{{x_n}}}R\left( {\left\{ {x_n^{{\rm{Initial}}}} \right\}_{n = 1}^2} \right)} \right|} }}\\
 =& \frac{{{{\log }_2}(1.5)}}{{5\frac{\pi }{{6\lambda \ln 2}}}} \approx 0.1548\;\lambda /{\rm{s}}.
\end{split}
\end{equation}

\textbf{Case~ii): $x_1^{\mathrm{Initial}}=5\lambda,\; x_2^{\mathrm{Initial}}=5.5\lambda$}. Here the initial spacing is $|\Delta_0|=0.5\lambda$. Following the same derivation as in Case~i) (details omitted for brevity), we obtain
\begin{equation}
\begin{split}{}
R\!\left(\{x_n^{\mathrm{Initial}}\}_{n=1}^2\right) = \log_2\!\left(1+\sin^2\frac{\pi}{16}\right)\;{\rm{bits/s/Hz}},
\end{split}
\end{equation}
\begin{equation}
\begin{split}{}
&\sum\nolimits_{n = 1}^2 {\left| {{\nabla _{{x_n}}}R\left( {\{ x_n^{{\rm{Initial}}}\} _{n = 1}^2} \right)} \right|}  \\
=& \frac{\pi }{{4\lambda \ln 2}}\frac{{\sin (\pi /8)}}{{1 + {{\sin }^2}(\pi /16)}},
\end{split}
\end{equation}
and consequently
\begin{equation}
\begin{split}{}
{V_{{\rm{th}}}} =& \frac{{{{\log }_2}\left( {1 + {{\sin }^2}\frac{\pi }{{16}}} \right)}}{{5\frac{\pi }{{4\lambda \ln 2}}\frac{{\sin (\pi /8)}}{{1 + {{\sin }^2}(\pi /16)}}}} \approx  0.02581\;\lambda /{\rm{s}}.
\end{split}
\end{equation}

After obtaining $V_{\mathrm{th}}$ for the two cases, we use the expressions in (26a) and (27a) to perform the one-dimensional search over $t_{\mathrm{mov}}$ for each $V_{\max}$ (here we set ${V_{\max }} \in [0,\lambda ]$ since this range is enough to verify Proposition 1). Given ${V_{\max }}$, the optimal movement duration that maximizes the effective throughput is denoted as $t_{{\rm{mov}}}^*({V_{\max }})$. Fig. 3 plots $t_{{\rm{mov}}}^*({V_{\max }})$ versus $V_{\max}$ for both cases, from which we can clearly verify the correctness of Proposition 1. Specifically, for Case i), when $V_{\max}\le 0.1548\lambda/\text{s}$, the optimal movement duration is exactly $t_{\mathrm{mov}}^*=0$; when $V_{\max}>0.1548\lambda/\text{s}$, $t_{\mathrm{mov}}^*$ becomes positive. The same behavior is observed for Case ii): $t_{\mathrm{mov}}^*=0$ for $V_{\max}\le 0.02581\lambda/\text{s}$, and $t_{\mathrm{mov}}^*>0$ for a larger speed. These observations perfectly match the threshold predicted by Proposition 1, confirming that the stationary strategy is optimal below $V_{\mathrm{th}}$ and that moving antennas becomes beneficial only when $V_{\max}$ exceeds the threshold.

\section{Simulation Results}
In this section, we conduct numerical simulations to validate the proposed schemes, where the optimized effective throughput with the general method and the fitting method are respectively denoted as ``OTGM'' and ``OTFM''. In addition, to perform comprehensive comparisons, we also consider three benchmarks:

i) \textbf{Upper bound}: The antenna movement speed is set to ${V_{\max }} \to \infty $, i.e., antennas instantaneously reach the optimized positions $\left\{ {{\bf{a}}_n^*} \right\}_{n = 1}^N$ and thus the resulted ideal throughput corresponding this scheme is $T{\log _2}\left( {1 + \gamma \left( {\left\{ {{\bf{a}}_n^*} \right\}_{n = 1}^N} \right)} \right)$;

ii) \textbf{Fixed movement duration with optimized antenna deployment (FMD+OAD)}: The antenna movement duration is fixed as ${t_{\rm{mov}}} = 0.2T$ and then antenna positions are carefully optimized based on the proposed general method;

iii) \textbf{Static deployment}: The antennas remain fixed throughout the entire interval, i.e., data transmission occurs over the whole period $T$ using the initial antenna positions.

Unless otherwise specified, the large-scale fading coefficient is defined as ${\beta _k} = {\beta _0}d_k^{ - {\alpha _0}}$, where ${\beta _0} = {10^{ - 4}}$ is the reference average channel power gain at 1 m and ${\alpha _0} = 2$ is the path loss exponent. The distance between the BS and user $k$ is set as ${d_k} = {10^2}$ m, $\forall k = 1,...,K$. The number of antennas at the BS is $N = 5$, the number of users is $K = 4$ with $\left\{ {{\theta _k}} \right\}_{k = 1}^K = \left\{ {\pi /2,\pi /4,\pi /6,\pi /8} \right\}$ and $\left\{ {{\phi _k}} \right\}_{k = 1}^K = \left\{ {\pi /3,\pi /5,\pi /7,\pi /8} \right\}$. The total time interval is $T = 8$ s, the side length of the square moving region is $L = 10\lambda $, the total transmit power budget of the BS is ${P_{\rm{tot}}} = 15$ dBm, the minimum distance between any two distinct antennas at the BS for avoiding mutual coupling effects is ${d_{\min }} = 0.5\lambda $, and the noise power is ${\sigma ^2} =  - 80$ dBm. The initial antenna deployment is set as ${{\bf{A}}^{{\rm{Initial}}}} = \left[ {\begin{array}{*{20}{c}}
{4.5}&5&{5.5}&6&{6.5}\\
0&0&0&0&0
\end{array}} \right]\lambda $.

Fig.~4(a) shows the effective throughput of OTGM versus $t_{\mathrm{mov}}$ for different $V_{\max}$. Two observations can be made: i) For each $V_{\max}$, the throughput first increases and then decreases, revealing the tradeoff between rate improvement and transmission time loss; ii) As $V_{\max}$ grows, the optimized $t_{\mathrm{mov}}$ shortens and the peak throughput rises, indicating that faster movement reduces the time penalty and enhances the utilization of spatial DoFs. Fig.~4(b) compares different schemes versus $V_{\max}$, from which we can observe that OTGM achieves the best performance, while OTFM closely approaches it with much lower complexity. Both outperform the benchmarks, where static deployment suffers from poor initial channel conditions, and FMD+OAD uses a fixed $t_{\mathrm{mov}}$ that fails to adapt to $V_{\max}$, causing either insufficient repositioning or unnecessary transmission time loss. As $V_{\max}$ increases, OTGM and OTFM converge to the upper bound.

 \begin{figure}
\centering
\includegraphics[width=7.5cm]{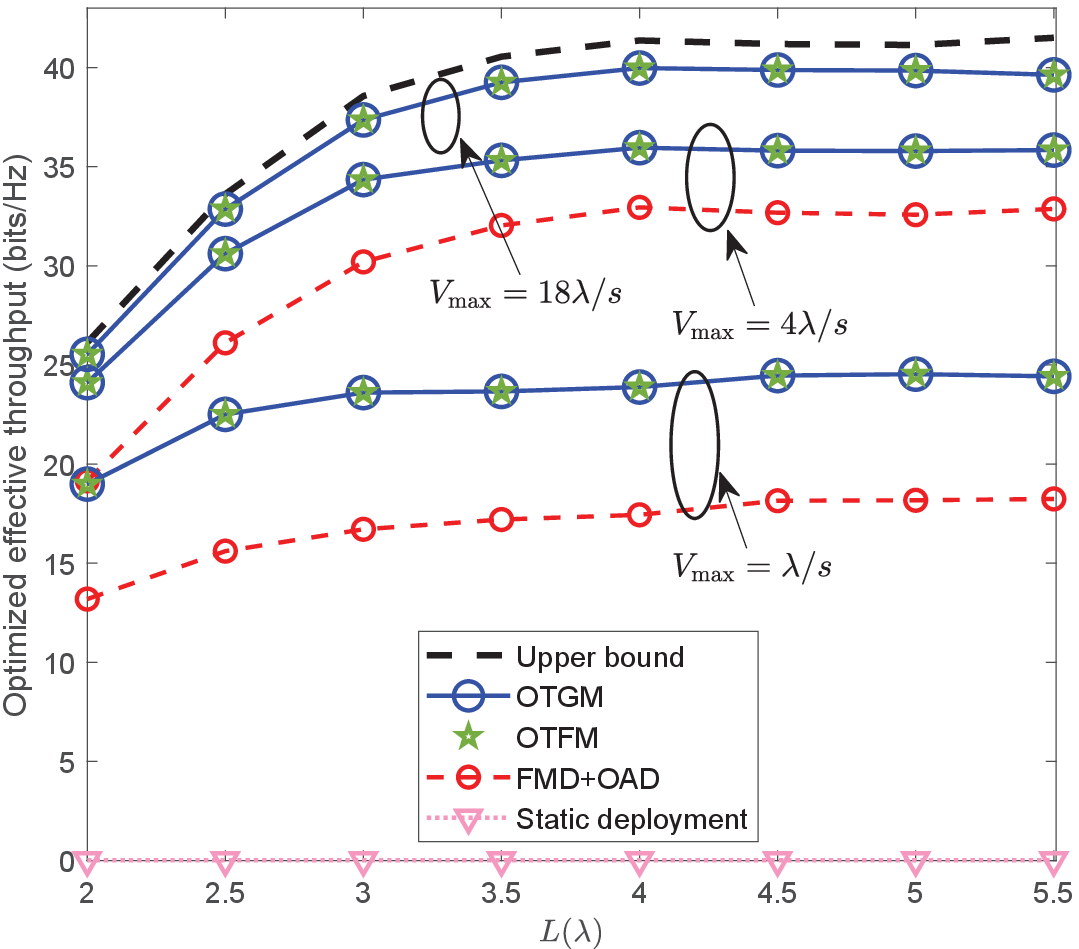}
\captionsetup{font=small}
\caption{Optimized effective throughput w.r.t. the length of the region's side $L$ under different schemes.} \label{fig:Fig1}
\vspace{-10pt}
\end{figure}

Fig. 5 illustrates the optimized effective throughput as a function of the movable region side length $L$ under different maximum speed limits, from which we can observe that: i) For any given $V_{\max}$, the throughput achieved by both OTGM and OTFM increases monotonically with $L$ until saturation. The threshold value of $L$ at which saturation occurs is determined exclusively by the optimized antenna positions without constraints in (13b) and is independent of $V_{\max}$; ii) In contrast, the saturation level itself strongly depends on $V_{\max}$: A higher movement speed consistently yields a higher asymptotic throughput. Notably, only when $V_{\max}$ is sufficiently large (e.g., $18\lambda/\text{s}$) does the throughput approach the theoretical upper bound; for a lower speed, the time overhead required for repositioning leads to a lower saturation ceiling; iii) Among the evaluated schemes, OTGM consistently achieves the highest effective throughput, while OTFM closely approaches this performance with substantially reduced computational complexity. The static deployment remains constant across all $L$ and $V_{\max}$, whereas the FMD+OAD scheme exhibits only marginal gains due to its fixed movement duration. Collectively, these results indicate that a moderate spatial region suffices to capture most of the attainable throughput improvement, provided the antenna movement duration and speed are jointly optimized.

 \begin{figure}
\centering
\includegraphics[width=7.5cm]{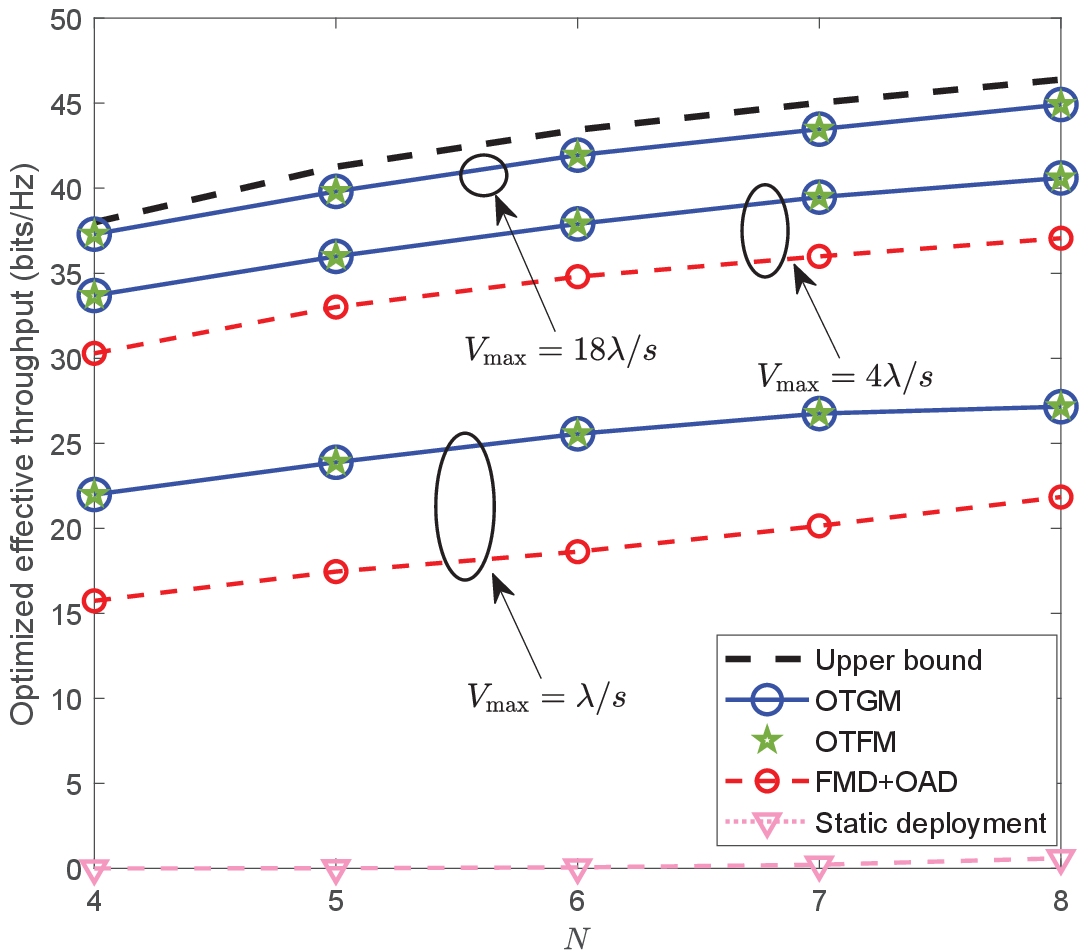}
\captionsetup{font=small}
\caption{Optimized effective throughput w.r.t. the number of antennas $N$ at the BS under different schemes.} \label{fig:Fig1}
\vspace{-10pt}
\end{figure}

Fig.~6 shows the optimized effective throughput versus the number of antennas $N$ under different maximum speed limits, where for each $N$, the initial antenna positions are uniformly placed along the $x$-axis with $y = 0$, centered within the movable region of side length $L = 4\lambda$, and with a fixed adjacent spacing of $0.5\lambda$. From Fig. 6 we can observe that: i) For a given $V_{\max}$, the throughput of all schemes increases with $N$ due to the higher DoFs and array gains, but the growth slows down as $N$ becomes large. The upper bound achieves the highest throughput, followed closely by OTGM and OTFM, while FMD+OAD and static deployment perform much worse; ii) The movement speed significantly affects performance. When $V_{\max}=18\lambda/s$, OTGM and OTFM approach the upper bound closely; when $V_{\max}$ reduces, the gap is much larger and throughput saturates earlier at a lower level, indicating that limited speed restricts the benefit of adding more antennas.

\begin{figure*}
    \centering
    \begin{minipage}{0.48\linewidth}
        \centering
        \includegraphics[width=7.1cm]{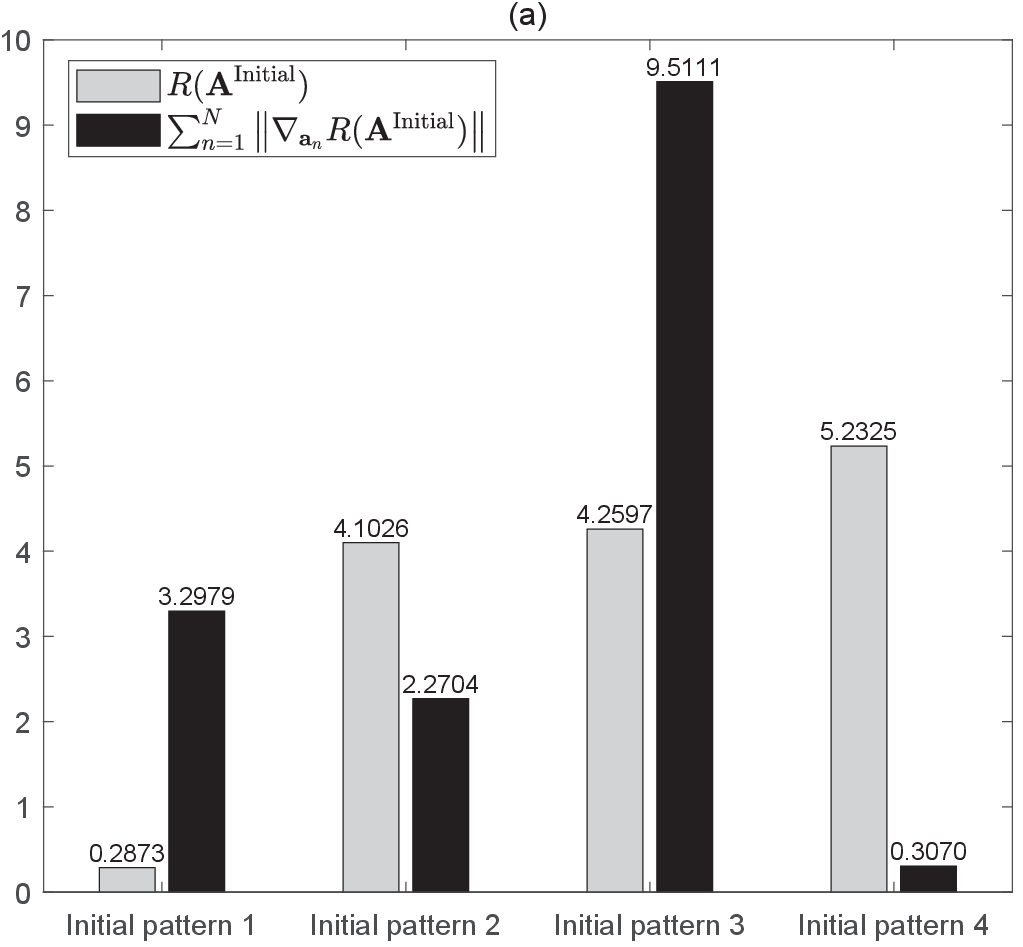}
    \end{minipage}
    \hfill
    \begin{minipage}{0.48\linewidth}
        \centering
        \includegraphics[width=7.5cm]{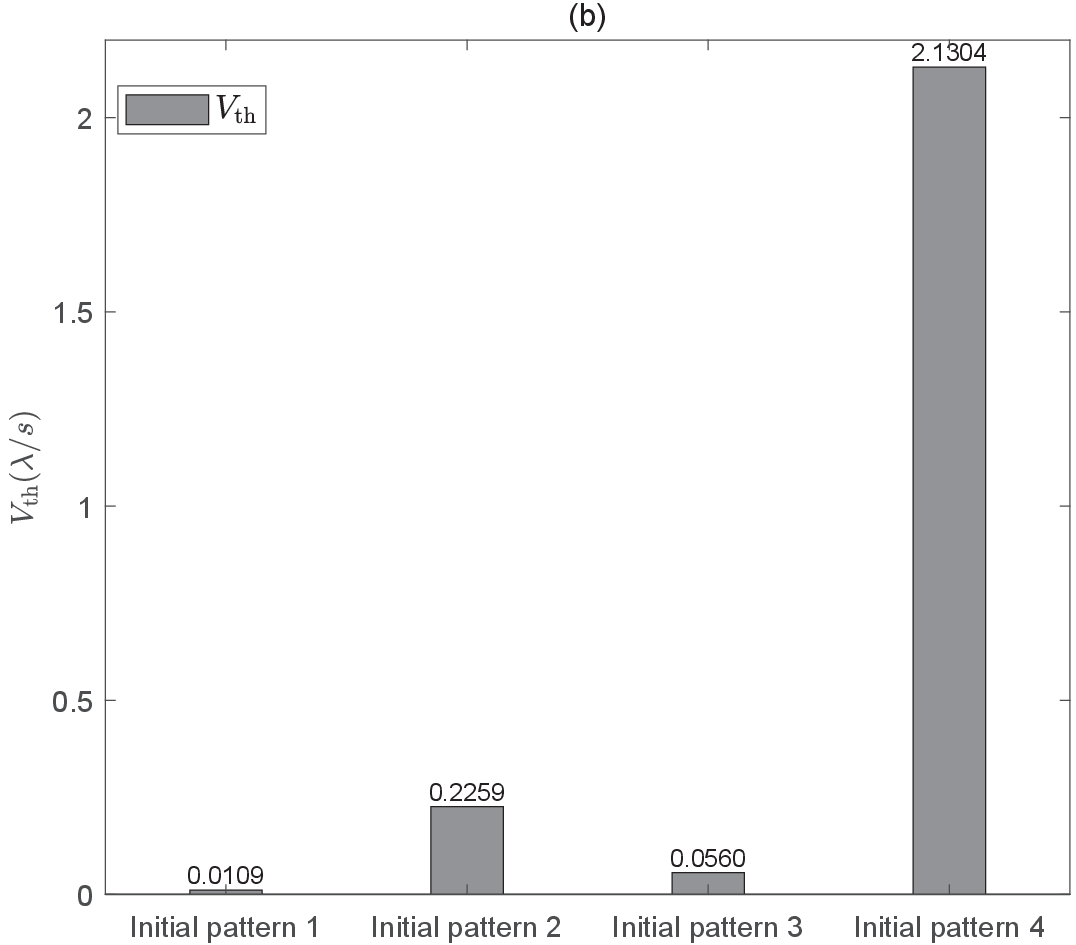}
    \end{minipage}
    \captionsetup{font=small}
    \caption{(a) ${R({{\bf{A}}^{{\rm{Initial}}}})}$ and $\sum\nolimits_{n = 1}^N {\left\| {{\nabla _{{{\bf{a}}_n}}}R({{\bf{A}}^{{\rm{Initial}}}})} \right\|} $ w.r.t. different initial antenna patterns; (b) ${V_{{\rm{th}}}}$ w.r.t. different initial antenna patterns.}
    \label{fig:2}
    \vspace{-1pt}
\end{figure*}

Fig.~7(a) depicts the initial rate \(R(\mathbf{A}^{\mathrm{Initial}})\) and the sum of gradient norms \(\sum_{n=1}^{N}\|\nabla_{\mathbf{a}_n}R(\mathbf{A}^{\mathrm{Initial}})\|\) for the four initial deployment patterns, i.e.,
\begin{equation} \nonumber
\begin{split}{}
{\bf{A}}_1^{{\rm{Initial}}} =& \left[ {\begin{array}{*{20}{c}}
1&{0.6}&{1.2}&{1.7}&{2.3}\\
1&1&{0.5}&{0.4}&0
\end{array}} \right]\lambda, \\
{\bf{A}}_2^{{\rm{Initial}}} =& \left[ {\begin{array}{*{20}{c}}
1&3&6&7&8\\
0&3&0&0&0
\end{array}} \right]\lambda, \\
{\bf{A}}_3^{{\rm{Initial}}} =& \left[ {\begin{array}{*{20}{c}}
1&1&6&6&3\\
0&5&0&5&3
\end{array}} \right]\lambda, \\
{\bf{A}}_4^{{\rm{Initial}}} =& \left[ {\begin{array}{*{20}{c}}
{3.9}&{0.7}&7&{4.5}&{8.7}\\
{2.25}&{9.5}&{3.35}&6&{0.45}
\end{array}} \right]\lambda,
\end{split}
\end{equation}
while Fig.~7(b) shows the corresponding theoretical speed threshold \(V_{\mathrm{th}}\) derived from Proposition~1.

The following trends emerge from these specific configurations: i) Pattern 1 exhibits a tightly clustered geometry where all antennas are confined within a small spatial region. This severe spatial concentration limits the available DoFs and results in a low initial rate $R(\mathbf{A}^{\mathrm{Initial}})$. However, the steep local gradient of the rate function in this highly coupled regime leads to a large sum of gradient norms, indicating substantial potential for channel enhancement through minor displacements. Consequently, the corresponding speed threshold $V_{\mathrm{th}}$ is the smallest among all patterns, implying that even a relatively slow movement speed suffices to justify the time overhead of antenna repositioning; ii) Patterns 2 and 3 represent more distributed configurations, with moderate antenna separations either predominantly along one dimension (Pattern~2) or across two dimensions with partial clustering (Pattern~3). These arrangements partially mitigate inter-user correlation, yielding moderate initial rate that exceeds that of Pattern~1. Simultaneously, the gradient sum decreases as the antennas reside in flatter regions of the rate surface. Accordingly, the resulting $V_{\mathrm{th}}$ values lie between the extremes, requiring a higher movement speed than Pattern~1 to warrant repositioning; iii) Pattern 4 approximates an optimal sparse deployment, with antennas well dispersed across the entire two-dimensional moving region. This geometry nearly achieves the minimum trace objective and thus attains the highest initial rate among all patterns. Moreover, being situated near a local maximum of the rate function, the gradient norms become vanishingly small. As a direct consequence, Pattern~4 yields the largest $V_{\mathrm{th}}$, signifying that only an exceptionally fast movement speed can overcome the loss of transmission time and render antenna repositioning beneficial.

\section{Conclusion}
This paper has revealed a fundamental tradeoff in MAs-aided multiuser downlink systems under movement delay: A longer movement duration improves channel conditions but reduces transmission time. To maximize the minimum effective throughput, we jointly optimized movement duration, antenna deployment, power allocations and beamforming. Using ZF-based beamforming and equal-rate power allocations, the problem was simplified to depend only on movement duration and antenna positions, for solving which we proposed a general method (one-dimensional search with penalty-based alternating optimization) and a low-complexity fitting method. A two-antenna two-user special case validated the tradeoff and the fitting models. We further derived a closed-form condition on the movement speed and the total time interval for which staying stationary is optimal. Simulations showed that the proposed schemes significantly outperform static and fixed-movement benchmarks and approach the upper bound at the high antenna movement speed.
\begin{appendix}
\end{appendix}
\section{The Closed-Form $\nabla_{\mathbf{a}_n^{t_{\mathrm{mov}}}} f(\mathbf{A}^{t_{\mathrm{mov}},(i-1)})$}
Note that
\begin{equation}
\setcounter{equation}{46}
\begin{split}
&{\nabla _{{\bf{a}}_n^{{t_{{\rm{mov}}}}}}}f({{\bf{A}}^{{t_{{\rm{mov}}}},(i - 1)}}) \\
=& {\left. {{\nabla _{{\bf{a}}_n^{{t_{{\rm{mov}}}}}}}f({{\bf{A}}^{{t_{{\rm{mov}}}}}})} \right|_{{{\bf{A}}^{{t_{{\rm{mov}}}}}} = {{\bf{A}}^{{t_{{\rm{mov}}}},(i - 1)}}}},
\end{split}
\end{equation}
where ${\nabla _{{\bf{a}}_n^{{t_{{\rm{mov}}}}}}}f({{\bf{A}}^{{t_{{\rm{mov}}}}}}) = {\left[ {\partial f/\partial {x_n},\partial f/\partial {y_n}} \right]^T}$, $f \buildrel \Delta \over = f({{\bf{A}}^{{t_{{\rm{mov}}}}}})$ and ${\bf{a}}_n^{{t_{{\rm{mov}}}}} \buildrel \Delta \over = {[{x_n},{y_n}]^T}$ for simplification. Then, it can be first derived that
\begin{equation}
\begin{split}
\partial f/\partial {x_n} =  - {\rm{tr}}\left( {{{\bf{G}}^{ - 1}}\left( {\frac{{\partial {{\bf{H}}^H}}}{{\partial {x_n}}}{\bf{H}} + {{\bf{H}}^H}\frac{{\partial {\bf{H}}}}{{\partial {x_n}}}} \right){{\bf{G}}^{ - 1}}} \right),
\end{split}
\end{equation}
where ${\bf{H}} \buildrel \Delta \over = {\bf{H}}({{\bf{A}}^{{t_{{\rm{mov}}}}}})$ and ${\bf{G}} = {{\bf{H}}^H}{\bf{H}}$. Subsequently, note the matrix $\frac{\partial \mathbf{H}}{\partial x_n}$ has zeros everywhere except its $n$-th row, which is
\begin{equation}
\begin{split}
{\left( {\frac{{\partial {\bf{H}}}}{{\partial {x_n}}}} \right)_{n,:}} = \left[ {\frac{{\partial {{[{\bf{H}}]}_{n,1}}}}{{\partial {x_n}}},...,\frac{{\partial {{[{\bf{H}}]}_{n,K}}}}{{\partial {x_n}}}} \right].
\end{split}
\end{equation}
By defining ${{\bf{b}}_k} = {[\cos {\theta _k}\sin {\phi _k},\sin {\theta _k}]^T} = {[{b_{k,x}},{b_{k,y}}]^T}$ and differentiating $[\mathbf{H}]_{n,k}$ ($k = 1,...,K$) w.r.t. $x_n$ gives
\begin{equation}
\begin{split}
\frac{\partial [\mathbf{H}]_{n,k}}{\partial x_n} =& \sqrt{\beta_k}\, e^{-j\frac{2\pi}{\lambda} (x_n b_{k,x} + y_n b_{k,y})} \cdot \left(-j\frac{2\pi}{\lambda} b_{k,x}\right) \\
=& -j\frac{2\pi}{\lambda} b_{k,x} [\mathbf{H}]_{n,k}.
\end{split}
\end{equation}

In addition, note that $\frac{{\partial {{\bf{H}}^H}}}{{\partial {x_n}}} = {\left( {\frac{{\partial {\bf{H}}}}{{\partial {x_n}}}} \right)^H}$, which has only the $n$-th column non-zero:
\begin{equation}
\begin{split}
{\left( {\frac{{\partial {{\bf{H}}^H}}}{{\partial {x_n}}}} \right)_{:,n}} =& {\left[ {\frac{{\partial {{[{{\bf{H}}^H}]}_{1,n}}}}{{\partial {x_n}}},...,\frac{{\partial {{[{{\bf{H}}^H}]}_{K,n}}}}{{\partial {x_n}}}} \right]^T}\\
 =& {\left[ {j\frac{{2\pi }}{\lambda }{b_{1,x}}[{\bf{H}}]_{n,1}^*,...,j\frac{{2\pi }}{\lambda }{b_{K,x}}[{\bf{H}}]_{n,K}^*} \right]^T}.
\end{split}
\end{equation}

Based on (47), (48), (49) and (50), and using the cyclic property of the trace together with the sparsity pattern, we obtain after simplification
\begin{equation}
\begin{split}
\partial f/\partial x_n = -\frac{4\pi}{\lambda} \sum_{k=1}^{K} b_{k,x} \operatorname{Im}\Bigl( \bigl[ \mathbf{G}^{-2} \mathbf{H}^H \bigr]_{k,n} [\mathbf{H}]_{n,k} \Bigr),
\end{split}
\end{equation}
where $\mathbf{G}^{-2} = \mathbf{G}^{-1}\mathbf{G}^{-1}$.

Similarly, the partial derivative w.r.t. $y_n$ is
\begin{equation}
\begin{split}
\partial f/\partial y_n = - \operatorname{tr}\left( \mathbf{G}^{-1} \left( \frac{\partial \mathbf{H}^H}{\partial y_n} \mathbf{H} + \mathbf{H}^H \frac{\partial \mathbf{H}}{\partial y_n} \right) \mathbf{G}^{-1} \right),
\end{split}
\end{equation}
and following the same steps yields
\begin{equation}
\begin{split}
\partial f/\partial y_n = -\frac{4\pi}{\lambda} \sum_{k=1}^{K} b_{k,y} \operatorname{Im}\Bigl( \bigl[ \mathbf{G}^{-2} \mathbf{H}^H \bigr]_{k,n} [\mathbf{H}]_{n,k} \Bigr).
\end{split}
\end{equation}

Combining the two components, we obtain the gradient vector
\begin{equation}
\begin{split}
{\nabla _{{\bf{a}}_n^{{t_{{\rm{mov}}}}}}}f({{\bf{A}}^{{t_{{\rm{mov}}}}}}) &=  - \frac{{4\pi }}{\lambda }\\
 \times & \sum\nolimits_{k = 1}^K {{{\bf{b}}_k}{\mathop{\rm Im}\nolimits} \left( {{{[{{\bf{G}}^{ - 2}}{{\bf{H}}^H}]}_{k,n}}{\mkern 1mu} {{[{\bf{H}}]}_{n,k}}} \right)}.
\end{split}
\end{equation}

Finally, for the $(i-1)$-th iteration, the gradient is evaluated at $\mathbf{A}^{t_{\mathrm{mov}},(i-1)}$ as in (46). This closed-form expression is directly used in the projected gradient descent update in (16).

\normalem
\bibliographystyle{IEEEtran}
\bibliography{IEEEabrv,mybib}

\begin{thebibliography}{10}
\providecommand{\url}[1]{#1}
\csname url@samestyle\endcsname
\providecommand{\newblock}{\relax}
\providecommand{\bibinfo}[2]{#2}
\providecommand{\BIBentrySTDinterwordspacing}{\spaceskip=0pt\relax}
\providecommand{\BIBentryALTinterwordstretchfactor}{4}
\providecommand{\BIBentryALTinterwordspacing}{\spaceskip=\fontdimen2\font plus
\BIBentryALTinterwordstretchfactor\fontdimen3\font minus
  \fontdimen4\font\relax}
\providecommand{\BIBforeignlanguage}[2]{{%
\expandafter\ifx\csname l@#1\endcsname\relax
\typeout{** WARNING: IEEEtran.bst: No hyphenation pattern has been}%
\typeout{** loaded for the language `#1'. Using the pattern for}%
\typeout{** the default language instead.}%
\else
\language=\csname l@#1\endcsname
\fi
#2}}
\providecommand{\BIBdecl}{\relax}
\BIBdecl

\bibitem{MIMO_1}
A.~PAULRAJ, D.~GORE, R.~NABAR, and H.~BOLCSKEI, ``An overview of {MIMO}
  communications - a key to gigabit wireless,'' \emph{Proc. IEEE}, vol.~92,
  no.~2, pp. 198--218, 2004.

\bibitem{MIMO_2}
E.~G. Larsson, O.~Edfors, F.~Tufvesson, and T.~L. Marzetta, ``Massive {MIMO}
  for next generation wireless systems,'' \emph{IEEE Commun. Mag.}, vol.~52,
  no.~2, pp. 186--195, 2014.

\bibitem{MIMO_3}
L.~Lu, G.~Y. Li, A.~L. Swindlehurst, A.~Ashikhmin, and R.~Zhang, ``An overview
  of massive {MIMO}: Benefits and challenges,'' \emph{IEEE J. Sel. Topics
  Signal Process.}, vol.~8, no.~5, pp. 742--758, 2014.

\bibitem{Lipeng1}
L.~Zhu, W.~Ma, and R.~Zhang, ``Modeling and performance analysis for movable
  antenna enabled wireless communications,'' \emph{IEEE Trans. Wireless
  Commun.}, vol.~23, no.~6, pp. 6234--6250, 2024.

\bibitem{KKWong1}
K.-K. Wong, A.~Shojaeifard, K.-F. Tong, and Y.~Zhang, ``Fluid antenna
  systems,'' \emph{IEEE Trans. Wireless Commun.}, vol.~20, no.~3, pp.
  1950--1962, 2021.

\bibitem{Lipeng2}
L.~Zhu, W.~Ma, and R.~Zhang, ``Movable antennas for wireless communication:
  Opportunities and challenges,'' \emph{IEEE Commun. Mag.}, vol.~62, no.~6, pp.
  114--120, 2024.

\bibitem{Xiaodan1}
X.~Shao and R.~Zhang, ``{6DMA} enhanced wireless network with flexible antenna
  position and rotation: Opportunities and challenges,'' \emph{IEEE Commun.
  Mag.}, vol.~63, no.~4, pp. 121--128, 2025.

\bibitem{Ningboyu}
B.~Ning, S.~Yang, Y.~Wu, P.~Wang, W.~Mei, C.~Yuen, and E.~Bj\"ornson, ``Movable
  antenna-enhanced wireless communications: General architectures and
  implementation methods,'' \emph{IEEE Wireless Commun.}, vol.~32, no.~5, pp.
  108--116, 2025.

\bibitem{Weidong}
D.~Wang, W.~Mei, B.~Ning, Z.~Chen, and R.~Zhang, ``Movable antenna enhanced
  wide-beam coverage: Joint antenna position and beamforming optimization,''
  \emph{IEEE Trans. Wireless Commun.}, vol.~25, pp. 3541--3558, 2026.

\bibitem{Xiaodan2}
X.~Shao, R.~Zhang, Q.~Jiang, and R.~Schober, ``6{D} movable antenna enhanced
  wireless network via discrete position and rotation optimization,''
  \emph{IEEE J. Sel. Areas Commun.}, vol.~43, no.~3, pp. 674--687, 2025.

\bibitem{Lipeng3}
L.~Zhu, W.~Ma, Z.~Xiao, and R.~Zhang, ``Performance analysis and optimization
  for movable antenna aided wideband communications,'' \emph{IEEE Trans.
  Wireless Commun.}, vol.~23, no.~12, pp. 18\,653--18\,668, 2024.

\bibitem{Lipeng4}
L.~Zhu, W.~Ma, W.~Mei, Y.~Zeng, Q.~Wu, B.~Ning, Z.~Xiao, X.~Shao, J.~Zhang, and
  R.~Zhang, ``A tutorial on movable antennas for wireless networks,''
  \emph{IEEE Commun. Surv. Tut.}, vol.~28, pp. 3002--3054, 2026.

\bibitem{Lipeng_CL}
L.~Zhu, W.~Ma, and R.~Zhang, ``Movable-antenna array enhanced beamforming:
  Achieving full array gain with null steering,'' \emph{IEEE Commun. Lett.},
  vol.~27, no.~12, pp. 3340--3344, 2023.

\bibitem{Wenyan1}
W.~Ma, L.~Zhu, and R.~Zhang, ``{MIMO} capacity characterization for movable
  antenna systems,'' \emph{IEEE Trans. Wireless Commun.}, vol.~23, no.~4, pp.
  3392--3407, 2024.

\bibitem{Xintai1}
X.~Chen, B.~Feng, Y.~Wu, X.-G. Xia, and C.~Xiao, ``Energy efficiency
  maximization for movable antenna-enhanced {MIMO} downlink system based on
  s-{CSI},'' \emph{IEEE Trans. Wireless Commun.}, vol.~25, pp. 4642--4657,
  2026.

\bibitem{Lipeng5}
L.~Zhu, W.~Ma, Z.~Xiao, and R.~Zhang, ``Movable antenna enabled near-field
  communications: Channel modeling and performance optimization,'' \emph{IEEE
  Trans. Commun.}, vol.~73, no.~9, pp. 7240--7256, 2025.

\bibitem{Xiandan3}
X.~Shao, Q.~Jiang, and R.~Zhang, ``{6D} movable antenna based on user
  distribution: Modeling and optimization,'' \emph{IEEE Trans. Wireless
  Commun.}, vol.~24, no.~1, pp. 355--370, 2025.

\bibitem{Xiandan4}
X.~Shao, R.~Zhang, Q.~Jiang, and R.~Schober, ``{6D} movable antenna enhanced
  wireless network via discrete position and rotation optimization,''
  \emph{IEEE J. Sel. Areas Commun.}, vol.~43, no.~3, pp. 674--687, 2025.

\bibitem{Guojie1}
G.~Hu, Q.~Wu, K.~Xu, J.~Ouyang, J.~Si, Y.~Cai, and N.~Al-Dhahir, ``Fluid
  antennas-enabled multiuser uplink: A low-complexity gradient descent for
  total transmit power minimization,'' \emph{IEEE Commun. Lett.}, vol.~28,
  no.~3, pp. 602--606, 2024.

\bibitem{Zhenyu1}
Z.~Xiao, X.~Pi, L.~Zhu, X.-G. Xia, and R.~Zhang, ``Multiuser communications
  with movable-antenna base station: Joint antenna positioning, receive
  combining, and power control,'' \emph{IEEE Trans. Wireless Commun.}, vol.~23,
  no.~12, pp. 19\,744--19\,759, 2024.

\bibitem{Guojie2}
G.~Hu, Q.~Wu, G.~Li, D.~Xu, K.~Xu, J.~Si, Y.~Cai, and N.~Al-Dhahir,
  ``Two-timescale design for movable antenna array-enabled multiuser uplink
  communications,'' \emph{IEEE Trans. Veh. Technol.}, vol.~74, no.~3, pp.
  5152--5157, 2025.

\bibitem{Lipeng6}
L.~Zhu, W.~Ma, B.~Ning, and R.~Zhang, ``Movable-antenna enhanced multiuser
  communication via antenna position optimization,'' \emph{IEEE Trans. Wireless
  Commun.}, vol.~23, no.~7, pp. 7214--7229, 2024.

\bibitem{Yifei1}
Y.~Wu, D.~Xu, D.~Wing Kwan~Ng, W.~Gerstacker, and R.~Schober, ``Globally
  optimal movable antenna-enabled multiuser communication: Discrete antenna
  positioning, power consumption, and imperfect {CSI},'' \emph{IEEE Trans.
  Commun.}, vol.~73, no.~10, pp. 9903--9923, 2025.

\bibitem{Songjie1}
S.~Yang, W.~Lyu, B.~Ning, Z.~Zhang, and C.~Yuen, ``Flexible precoding for
  multi-user movable antenna communications,'' \emph{IEEE Wireless Commun.
  Lett.}, vol.~13, no.~5, pp. 1404--1408, 2024.

\bibitem{Ziyuan1}
Z.~Zheng, Q.~Wu, W.~Chen, and G.~Hu, ``Two-timescale design for movable
  antenna-enabled multiuser {MIMO} systems,'' \emph{IEEE Trans. Commun.},
  vol.~73, no.~11, pp. 10\,554--10\,571, 2025.

\bibitem{Wenyan_CE1}
Z.~Xiao, S.~Cao, L.~Zhu, Y.~Liu, B.~Ning, X.-G. Xia, and R.~Zhang, ``Channel
  estimation for movable antenna communication systems: A framework based on
  compressed sensing,'' \emph{IEEE Trans. Wireless Commun.}, vol.~23, no.~9,
  pp. 11\,814--11\,830, 2024.

\bibitem{Wenyan_CE2}
W.~Ma, L.~Zhu, and R.~Zhang, ``Compressed sensing based channel estimation for
  movable antenna communications,'' \emph{IEEE Commun. Lett.}, vol.~27, no.~10,
  pp. 2747--2751, 2023.

\bibitem{Wenyan_TSP}
------, ``Movable antenna enhanced integrated sensing and communication via
  antenna position optimization,'' \emph{IEEE Trans. Signal Process.}, pp.
  1--17, 2026.

\bibitem{Wanting}
W.~Lyu, S.~Yang, Y.~Xiu, Z.~Zhang, C.~Assi, and C.~Yuen, ``Movable antenna
  enabled integrated sensing and communication,'' \emph{IEEE Trans. Wireless
  Commun.}, vol.~24, no.~4, pp. 2862--2875, 2025.

\bibitem{Wu_ISAC}
Q.~Wu, Z.~Zheng, Y.~Gao, W.~Mei, X.~Wei, W.~Chen, and B.~Ning, ``Integrating
  movable antennas and intelligent reflecting surfaces ({MA}-{IRS}):
  Fundamentals, practical solutions, and {ISAC},'' \emph{IEEE Wireless
  Commun.}, vol.~33, no.~1, pp. 155--163, 2026.

\bibitem{Weixin}
X.~Wei, W.~Mei, Q.~Wu, Q.~Jia, B.~Ning, Z.~Chen, and J.~Fang, ``Movable
  antennas meet intelligent reflecting surface: Friends or foes?'' \emph{IEEE
  Trans. Commun.}, vol.~73, no.~11, pp. 12\,756--12\,770, 2025.

\bibitem{Haoze}
H.~Wang, X.~Shao, B.~Zheng, X.~Shi, and R.~Zhang, ``Passive six-dimensional
  movable antenna ({6DMA})-assisted multiuser communication,'' \emph{IEEE
  Wireless Commun. Lett.}, vol.~14, no.~4, pp. 1014--1018, 2025.

\bibitem{Wu_MIS}
Z.~Zheng, Q.~Wu, W.~Chen, X.~Wu, and W.~Zhu, ``Movable intelligent surface
  ({MIS}) for wireless communications: Architecture, modeling, algorithm, and
  prototyping,'' \emph{IEEE Trans. Wireless Commun.}, vol.~25, pp. 5749--5765,
  2026.

\bibitem{MA_IRS}
Y.~Gao, Q.~Wu, Z.~Zheng, Y.~Zhu, W.~Chen, X.~Lin, and S.~Shen, ``Two-scale
  spatial deployment for cost-effective wireless networks via cooperative irss
  and movable antennas,'' \emph{arXiv preprint arXiv:2601.09463}, 2026.

\bibitem{MA_IRS1}
Q.~Wu, Z.~Zheng, Y.~Gao, W.~Mei, X.~Wei, W.~Chen, and B.~Ning, ``Integrating
  movable antennas and intelligent reflecting surfaces ({MA}-{IRS}):
  Fundamentals, practical solutions, and {ISAC},'' \emph{IEEE Wireless
  Commun.}, vol.~33, no.~1, pp. 155--163, 2026.

\bibitem{MA_IRS2}
Z.~Zheng, Q.~Wu, W.~Chen, W.~Zhu, and Y.~Gao, ``Movable intelligent
  surface-enabled wireless communications: When static phase shift meets
  mechanical reconfigurability,'' \emph{arXiv preprint arXiv:2511.17058}, 2025.

\bibitem{MA_IRS3}
Z.~Zheng, Q.~Wu, Y.~Zhu, W.~Chen, Y.~Gao, and H.~Wang, ``Wireless sensing with
  movable intelligent surface,'' \emph{IEEE J. Sel. Topics Signal Process.},
  pp. 1--16, 2026.

\bibitem{peng2025double}
Q.~Peng, Q.~Wu, G.~Chen, W.~Chen, S.~Shen, and S.~Ma, ``Cooperative rotatable
  {IRSs} for wireless communications: {Joint} beamforming and orientation
  optimization,'' \emph{arXiv preprint arXiv:2512.14037}, 2025.

\bibitem{peng2025single}
Q.~Peng, Q.~Wu, G.~Chen, W.~Chen, S.~Ma, S.~Shen, and R.~Zhang, ``Rotatable
  {IRS} aided wireless communication,'' \emph{arXiv preprint arXiv:2511.10006},
  2025.

\bibitem{Lipeng_Survey}
W.~Ma, L.~Zhu, Y.~Tan, B.~Zheng, Y.~Zhang, Y.~Zhang, K.~Ying, Z.~Gao, H.~Sun,
  X.~Shao, Z.~Xiao, D.~Niyato, and R.~Zhang, ``A survey on reconfigurable and
  movable antennas for wireless communications and sensing,'' \emph{IEEE
  Commun. Surv. Tut.}, vol.~28, pp. 4842--4882, 2026.

\bibitem{DZG}
Z.~Ding, R.~Schober, and H.~Vincent~Poor, ``Flexible-antenna systems: A
  pinching-antenna perspective,'' \emph{IEEE Trans. Commun.}, vol.~73, no.~10,
  pp. 9236--9253, 2025.

\bibitem{peng2025rotatable}
X.~Peng, Q.~Wu, Z.~Zheng, W.~Chen, Y.~Zhu, and Y.~Gao, ``Rotatable antenna
  enabled spectrum sharing: Joint antenna orientation and beamforming design,''
  \emph{arXiv preprint arXiv:2509.19912}, 2025.

\bibitem{Honghao}
H.~Wang, Q.~Wu, Y.~Gao, W.~Chen, W.~Mei, G.~Hu, and L.~Xu, ``Throughput
  maximization for movable antenna systems with movement delay consideration,''
  \emph{IEEE Trans. Wireless Commun.}, vol.~25, pp. 883--899, 2026.

\bibitem{Haiquan}
H.~Lu and Y.~Zeng, ``Near-field modeling and performance analysis for
  multi-user extremely large-scale {MIMO} communication,'' \emph{IEEE Commun.
  Lett.}, vol.~26, no.~2, pp. 277--281, 2022.

\bibitem{Penalty}
Y.~Jin, Q.~Lin, Y.~Li, H.~Zhu, B.~Cheng, Y.-C. Wu, and R.~Zhang, ``A general
  optimization framework for tackling distance constraints in movable
  antenna-aided systems,'' \emph{IEEE Trans. Wireless Commun.}, vol.~25, pp.
  10\,869--10\,885, 2026.

\end{thebibliography}

\end{document}